%% file: revised-main.tex
\documentclass[10pt,leqno]{amsart}

\usepackage[]{forest}
\forestset{.style={for tree=
{parent anchor=south, child anchor=north,align=center,inner sep=2pt}}}
\usepackage{graphicx}
\usepackage{indentfirst,csquotes}

\usepackage[english]{babel}
\usepackage[utf8x]{inputenc}
\usepackage[T1]{fontenc}
\usepackage{natbib}
\usepackage{subcaption}
\usepackage{mathtools, amsthm}
\usepackage{booktabs}
\usepackage{tabularx}
\usepackage{pifont}
\usepackage[a4paper,top=3cm,bottom=3cm,left=3cm,right=3cm,marginparwidth=1.75cm]{geometry}

\usepackage{amsmath}
\usepackage{amsfonts}
\usepackage{bm}
\usepackage[colorinlistoftodos]{todonotes}
\usepackage[colorlinks=true, allcolors=blue]{hyperref}
\usepackage{amssymb}
\usepackage{mathrsfs}
\usepackage{epstopdf}
\usepackage{mathtools}
\usepackage{enumitem}
\usepackage{multirow}
\usepackage{color,soul}
\usepackage{diagbox}

\newtheorem{thm}{Theorem}
\newtheorem{cor}{Corollary}

\usepackage[noend]{algpseudocode}
\usepackage{algorithm}
\newtheorem{remark}{Remark}
\newcommand{\ES}{\text{ES}}

\newcommand{\mcA}{\mathcal{A}}

\newcommand{\mcJ}{\mathcal{J}}
\newcommand{\mcL}{\mathcal{L}}
\newcommand{\mcR}{\mathcal{R}}

\newcommand{\mcO}{\mathcal{O}}
\newcommand{\mcS}{\mathcal{S}}
\newcommand{\mcQ}{\mathcal{Q}}

\newcommand{\mbE}{\mathbb{E}}
\newcommand{\mbR}{\mathbb{R}}
\newcommand{\mbI}{\mathbb{I}}
\newcommand{\mbP}{\mathbb{P}}

\newcommand{\Fhat}{\widehat{F}}

\newcommand{\ybar}{\bar{y}}
\newcommand{\inP}{\mbox{$\,\stackrel{\scriptsize{\mbox{p}}}{\rightarrow}\,$}}
\DeclareMathOperator*{\argmin}{arg\,min}

\newcommand \tcb{\color{black}}
\newcommand \tck{\color{black}}
\newcommand{\revision}[1]{\tcb #1\tck}
\def\hb#1{[BLINDED FOR REVIEW]}

\usepackage{xr}
\makeatletter

\newcommand*{\addFileDependency}[1]{
\typeout{(#1)}
\@addtofilelist{#1}
\IfFileExists{#1}{}{\typeout{No file #1.}}
}\makeatother

\newcommand*{\myexternaldocument}[1]{%
\externaldocument{#1}%
\addFileDependency{#1.tex}%
\addFileDependency{#1.aux}%
}

\myexternaldocument{revised-supp}

\begin{document}
\def\spacingset#1{\renewcommand{\baselinestretch}%
{#1}\small\normalsize} \spacingset{1}
%
 \title{Building Trees for Probabilistic Prediction via Scoring Rules}
 \author[Shashaaani, S\"{u}rer, Plumlee, and Guikema]{Sara Shashaani
    \\
 Department of Industrial and Systems Engineering\\
 North Carolina State University\vspace{.3cm}\\
 \"{O}zge S\"{u}rer \\ 
Department of Information Systems and Analytics \\\ Miami University\vspace{.3cm}\\
 Matthew Plumlee\\
 Principal Applied Scientist\\
 Amazon Inc.\vspace{.3cm}\\
 and
 Seth Guikema\\
 Department of Industrial and Operations Engineering\\
 University of Michigan\\
 }
 \maketitle


\begin{abstract}
Decision trees built with data remain in widespread use for nonparametric prediction. Predicting probability distributions is preferred over point predictions when uncertainty plays a prominent role in analysis and decision-making. We study modifying a tree to produce nonparametric predictive distributions. We find the standard method for building trees may not result in good predictive distributions and propose changing the splitting criteria for trees to one based on proper scoring rules. Analysis of both simulated data and several real datasets demonstrates that using these new splitting criteria results in trees with improved predictive properties considering the entire predictive distribution.
\end{abstract}


\spacingset{2} 

\section{Introduction}
Binary trees that partition continuous response variables based on predictor variables have been proven useful for nonparametric regression \citep{breiman1984classification}. \revision{Nonparametric regression is a general class of regression models that do not assume a parametric form for the relationship between predictors and dependent variables; binary trees can be considered an instance of them. } After the tree is statistically learned via training data, any new data point maps to a leaf (terminal node) in the tree based on the predictors' values. The resulting output for prediction is typically a statistic measuring the center of responses (in the training data) that belong to the same node \citep{hastie2009elements}. However, this statistic yields a decidedly deterministic forecast. In contrast, for many applications, such as weather and finance, it makes sense to predict probabilistically to communicate the stochastic nature of the system. 

\paragraph{\revision{Goal of prediction:}} A probabilistic prediction has two major goals: (i) to have the observations be consistent with the predictive distribution, and (ii) to concentrate (sharpen) the prediction as much as possible given the predictor variables \citep{gneiting2007probabilistic}. Thus, a reliable predictive distribution communicates both the magnitude of the prediction and the amount of uncertainty. The user could then convert the predictive distribution into a prediction. The best prediction might be a measure of center (mean), but it might be another feature of the predictive distribution based on the use case. For example, when predicting the number of power outages in a region after a storm, a $95\%$ upper bound would provide a picture of high-risk areas (quantile). In another scenario, one might wish to find the probability that an online article does not meet a view target (tail probability). In yet another scenario, predicting the variance of power consumption in a neighborhood can be critical to understanding potential load imbalance risks (second moment). In all of these examples, the nature of the prediction cannot be gleaned from the sample mean. 

\subsection{Probabilistic Predictions with Trees} Given a tree, one can generate a nonparametric predictive distribution for each terminal node using that node's empirical cumulative distribution function (ECDF). This suggestion by \cite{meinshausen2006quantile} is an input for the popular \emph{quantile regression forests}. But there is no guarantee that standard trees learned from data will have good predictive properties. 

The ``standard'' tree is built through a recursion where at each terminal node, potential splits of the tree are considered, and the split that most reduces the sum of squared errors (SSE) is chosen. Section \ref{sec:illus} of this article will demonstrate that even in simple conditions, trees built by splitting based on the SSE criteria do not necessarily possess good predictive properties. There have been other criteria designed for splitting rules beyond this typical approach. Splitting rules for \emph{classification} were well-dissected by \cite{taylor1993block} and \cite{breiman1996some}, which covered the Gini criteria and entropy. There appears to be less extensive literature on splitting rules for continuous prediction in nonparametric regression. 
Other splitting criteria such as log-rank \citep{leblanc1993survival}, likelihoods \citep{su2004maximum,zeileis2008model}, and treatment difference models \citep{su2009subgroup,athey2016recursive} are specialized and/or rely on parametric frameworks. \cite{athey2019generalized} offer a fully nonparametric method for predicting a quantity, not a predictive distribution. 

\subsection{Summary of Contributions and Insights}
This article offers novelty by suggesting splitting criteria for trees based on \emph{scoring rules}.  Scoring rules assess predictions and have a lengthy history in statistics, information theory, and convex analysis \citep{gneiting2007strictly}. \revision{For ease of exposition here, we use simple notations. } A scoring rule $S(F,y)$ takes a predictive distribution $F$ (throughout the paper this means the cumulative distribution function) and a realized quantity $y$ and converts it into a scalar score. We will consider negatively oriented scoring rules, where the smaller the score is, the better we have done. A \emph{proper} scoring rule encourages the predictor to provide the true distribution, which means one makes careful assessments and is honest about uncertainty \citep{garthwaite2005statistical}. Scoring rules tend to reward both goals of probabilistic prediction, though the respective importance is often hidden from the user. 

The novel splitting criteria are as follows. Consider splitting a current terminal node into two smaller terminal nodes with data \revision{$\{y_1,\ldots,y_l\}$ and $\{y_{l+1},\ldots,y_{l+r}\}$, to the left and right subsets respectively. Then we choose the split that minimizes
\begin{equation}
\sum_{i=1}^l S \left(\Fhat_{\mcL},y_i\right)+ \sum_{i=1}^r S \left(\Fhat_{\mcR},y_{l+i}\right), \label{eq:splittingrule}
\end{equation}
where $\Fhat_{\mcL}$ and $\Fhat_{\mcR}$ are the predictive ECDFs of $y$ }relating to the left and right side 
of the split. \revision{As discussed earlier, finding an optimal split in a standard tree is through minimizing SSE, which is itself a scoring rule. } Despite the relative simplicity of this formulation of splitting rules, the authors have found no reference to this mechanism with respect to building trees. The closest attempts in this direction have been studies on quantile-based loss functions~\citep{bhat2015towards}, density forecasts~\citep{iacopini2022proper}, or gradient forests~\citep{athey2019generalized}.

\revision{By considering scoring rules other than SSE, we aim to improve the predictability of trees. } But perhaps the most promising advantage of our proposed method is the application-dependent choice of a scoring rule. \revision{In different applications, probabilistic properties other than the mean behavior can be of importance. For example, interval scores that encourage narrow and consistent predictive intervals can be beneficial }if a user routinely uses only predictive intervals from the predictive distribution~\citep{christoffersen1998evaluating}, \revision{or in reliability applications and prediction of high-risk (extreme) events.  Two-moment scores are more useful when mean and variance are both of importance, or with datasets that possess significant heteroscedasticity. While the aforementioned scoring rules might need to be justified for the context, continuously ranked probabilistic scores (CRPS) are strictly proper scoring rules that are already understood as a better fit in weather forecasting  \citep{taillardat2016calibrated,vogel2018skill}. } 
See Section \ref{sec:background_score} for a modest background on these scoring rules and their computational costs. 

We describe the algorithmic structure for building score-based trees and pruning them in Section \ref{sec:algorithm} along with an important structural property of proper scoring rules: monotonic improvement. Asymptotic analysis of splits in Section~\ref{sec:empirical-split} provides a more general insight into scoring rules' necessary conditions for consistency. By examining synthetic and real datasets in Section \ref{sec:numerics}, we show that different scoring rules return substantially different trees in real, practical examples. \revision{ Our experiments confirm that when data is not completely summarized by the mean value, trees built with non-SSE scoring rules provide better predictions. Additionally, non-SSE trees can improve the SSE performance beyond traditional trees, and interval scores and CRPS achieve good prediction no matter what the goal of probabilistic prediction is. }
Section \ref{sec:conclude} closes the paper with some remarks on extensions to this approach.

\section{An illustration} \label{sec:illus}
This section will motivate the use of scoring rules to guide the splitting of tree models from a strictly statistical perspective. \revision{Throughout, we use script fonts for sets, the notation $[a]:=\{1,2,\ldots,a\}$ for some positive integer $a$, and $y_n\inP y$ for convergence in probability of a random sequence $y_n$ to a random variable $y$. Let $\{( x_i,y_i)\}_{i=1}^n$ be the available data with $x_i = (x_{i}^1,x_{i}^2,\ldots,x_{i}^p)$ representing the $p$ independent variables as potential predictors (features), and $y_i$ representing the (real-valued) response that we wish to predict for unseen data. We denote $\mcJ=\{1,2,\ldots,n\}$ as the index set of the whole data. We also use the notation $\Fhat_{\mcA}$ for the ECDF of data points whose index is in the index set $\mcA$. Specifically, $\Fhat_{\mcJ}(z) = \frac{1}{n} \sum_{i\in\mcJ} \mbI( y_{i}\leq z)$ is the empirical distribution of the entire training data. 
}

The SSE criterion cannot distinguish splits if the predictor variable $x$ impacts the distribution of the response variable $y$ but leaves the mean of $y$ unperturbed. Conversely, other scoring rules, which consider the entire distribution, can easily find these splits. To show this effect, we use an obvious shift of behavior in a small toy example where predictors $x_1,\ldots,x_{n}\sim\text{Unif}(-1,1)$ and the response variable is distributed as
\begin{equation}\small
y_i \sim\begin{cases}
\text{Normal} (\mu=1,\sigma=2) & \text{ if } x_i \in [-1,0],\\
\text{Exponential}(\lambda=1) &\text{ if } x_i \in (0,1].
\end{cases} \label{eq:illus_generation}
\end{equation}
Say we do not know the split occurs at $0$ but wish to build a tree of depth 1 to give good predictions.
\revision{For any split $(k,s)$, the traditional SSE score after split is 
\begin{align}
\text{SSE} (k,s) =& \sum_{i \in \mcL(k,s)} \underbrace{\left(y_i - \ybar_{\mcL(k,s)} \right)^2}_{S\left(\Fhat_{\mcL(k,s)},y_i\right)} + \sum_{i \in \mcR(k,s)} \underbrace{\left(y_i - \ybar_{\mcR(k,s)}\right)^2}_{S\left(\Fhat_{\mcR(k,s)},y_i\right)} \nonumber\\
=& \sum_{i=1}^{n} \left(y_i - \ybar_{\mcJ}   \right)^2 - |\mcL(k,s)| \left(\ybar_{\mcJ}- \ybar_{\mcL(k,s)}\right)^2 - |\mcR(k,s)| \left(\ybar_{\mcJ}- \ybar_{\mcR(k,s)}\right)^2 \nonumber\\
=& \sum_{i=1}^{n} \underbrace{\left(y_i - \ybar_{\mcJ}\right)^2}_{S\left(\Fhat_{\mcJ},y_i\right)} - \underbrace{\frac{|\mcL(k,s)||\mcR(k,s)|}{n} \left(\ybar_{\mcL(k,s)} - \ybar_{\mcR(k,s)}\right)^2}_{\text{reduction in SSE after split}},\label{eq:sse-split}
\end{align} which shows the SSE score will always reduce as a result of the split. In \eqref{eq:sse-split}, $(k,s)$ denotes the split using the $k$-th predictor based on its values less or greater than $s$. $\mcL(k,s) = \{i\in\mcJ: x_i^k \leq s\}$ and $\mcR(k,s) = \{i\in\mcJ: x_i^k > s\}$ denote the index subsets to the left and right of the  $(k,s)$ split, and  $\ybar_{\mcJ},\ybar_{\mcL(k,s)}$, and $\ybar_{\mcR(k,s)}$ denote sample mean of $y$'s whose indices are in $\mcJ,\mcL(k,s),$ and $\mcR(k,s)$. 
The optimal split is then given by
$(k^{\text{SSE}},s^{\text{SSE}}) = \argmin_{k\in[p],s\in\mbR} \text{SSE} (k,s)$.
Clearly, this criteria depends only on the sample means on each side of the split. If the sample means on each side are relatively close, SSE gives no information to guide the split; this criterion would likely be poor for splitting in our setting. 

We can alternatively use scoring rules such as CRPS with a simple implementation of 
\begin{equation}
\text{CRPS} (k,s) = \underbrace{\frac{1}{2 |\mcL(k,s)|} \sum_{i \in \mcL(k,s) } \sum_{j \in \mcL(k,s) } |y_i - y_j|}_{\sum_{i\in\mcL(k,s)}S\left(\Fhat_{\mcL(k,s)},y_i\right)} + \underbrace{\frac{1}{2 |\mcR(k,s)|} \sum_{i \in \mcR(k,s)} \sum_{j \in \mcR(k,s)}|y_i - y_j|}_{\sum_{i\in\mcR(k,s)}S\left(\Fhat_{\mcR(k,s)},y_i\right)}, \label{eq:crps_diff}
\end{equation}
 to guide the split, yielding 
$(k^{\text{CRPS}},s^{\text{CRPS}}) = \argmin_{k\in[p],s\in\mbR} \text{CRPS}(k,s)$. }
Compared to SSE, this criterion analyzes the difference between all of the values as opposed to just the sample means. We will discuss CRPS and its properties further in Section~\ref{sec:background_score}. See \citep{gneiting2007strictly} for a thorough description of CRPS. 

\begin{figure}[ht]
\centering
\includegraphics[width=.7\linewidth]{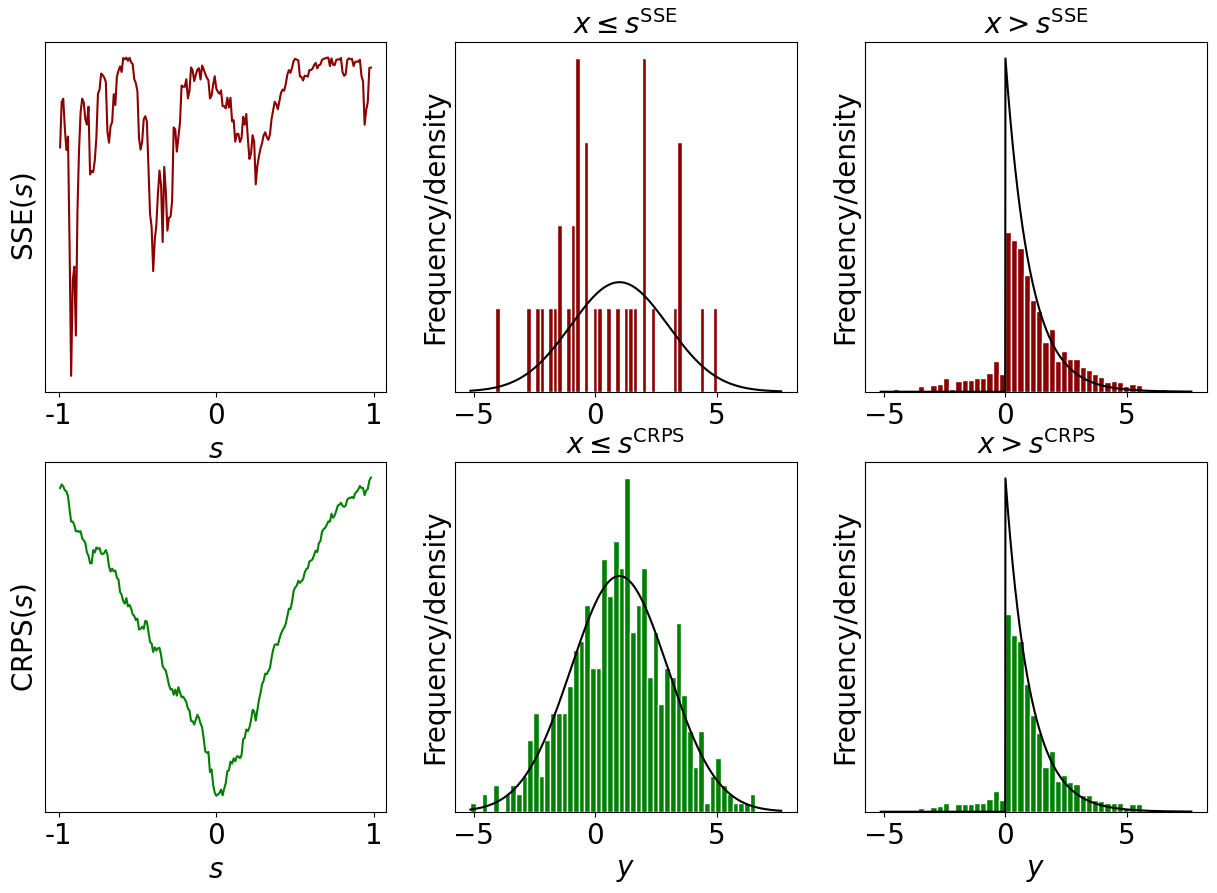}\caption{Splits resulting from the simulated experiment~\eqref{eq:illus_generation}. \revision{In this one-dimensional example, $k$ is always 1. } The left panels show the criteria versus the split point. The center (right) panels show the histogram from the data below (above) the selected split overlaying the true density when $x \in [-1,0]$ ($x \in (0,1]$).} \label{fig:ex1}
\end{figure}
Returning to our toy example, by changing $s$ in the $[-1,1]$ range, one can investigate where each of these criteria suggests the splitting must occur. Figure \ref{fig:ex1} illustrates the best split corresponding to each criterion and the density of data below and above that split value (overlaid by the known density) on a random dataset generated from (\ref{eq:illus_generation}) with $n = 1000$. The $\text{SSE}$ criterion is noisy, with no particular behavior where the true split is known. The $\text{CRPS}$ criterion also has some noise, but the trend focuses on the minimizer near $0$. In this experiment, constructing the tree by minimizing a proper score is superior to using the standard splitting criteria.

With the promise of this experiment, we propose the idea of constructing the tree by minimizing a criterion based on proper scoring rules. \revision{These criteria are used to find the split that minimizes the \emph{total score} $C(k,s)$: 
\begin{equation}
C(k,s):=\sum_{i\in\mcL(k,s)}S\left(\Fhat_{\mcL(k,s)},y_i\right)+\sum_{i\in\mcR(k,s)}S\left(\Fhat_{\mcR(k,s)},y_i\right).\label{eq:total-score}
\end{equation}}
While the proposed framework is broad, we work with several famous scores. 
Of course, the use of each score depends on the application, but the main point is to investigate whether contracting the trees with proper scores results in more reliable trees. 

\revision{\begin{remark}\label{rem:double-use}
    The training data is used for (a) constructing the ECDF and (b) calculating the scores. However, both of these steps are at the service of fitting a model to the data at hand, such that the fitted model can best mimic that data. Using separate sets of data or cross-validation for fitting, i.e., a training dataset to decide the splits and a separate validation dataset to calculate $\hat F$ for each split, will fail to minimize the score (loss) in the training dataset. 
    We note double purposes with the training data does not lead to overfit; in a traditional tree, too, the same data that provides a predictive distribution, is used to compute the score at the fitting step. To avoid overfitting, we control its root cause, i.e., model complexity~\citep{hastie2009elements}[Section 2.9], by carefully choosing the tree parameters and by a pruning mechanism via cross-validation; see Section~\ref{sec:parameters} and \ref{sec:pruning}.
\end{remark}}

\section{Background on Scoring Rules} \label{sec:background_score}
A scoring rule takes a distribution and an observed value and returns a score, often used to assess the closeness between the predictive distribution and reality. \cite{gneiting2007strictly}, \cite{dawid2007geometry}, and \cite{carvalho2016overview} provide reviews, summaries, and applications of scoring rules. Here we employ them to compare two or more alternative predictive distributions. However, there are other usages of scoring rules such as elicitation of distributions \citep*{garthwaite2005statistical}, missing value imputation \citep*{hasan2021missing}, and Bayesian utility theory \citep{bernardo2006bayesian}. 

Let $S(F,y)$ represent the score when distribution $F$ is used and $y$ is an observed continuous quantity, with $\ES(G,F) := \mbE_{y \sim F} S(G,y)$ representing the expected score of any distribution $G$ if the (true) generating distribution is $F$.
Scoring rules are negatively oriented, where smaller is better, and their most important property is propriety. A scoring rule is considered \emph{proper} if for all distributions $G$,
\begin{equation}
    \ES(F,F) \leq \ES(G,F),\label{eq:proper}
\end{equation}
and a scoring rule is strictly proper if for all $G \neq F$ the inequality is strict.
One interpretation of propriety is that if we choose a distribution to predict a quantity, the long-run average score is best minimized by selecting the true distribution.  The function $\ES(F,F)$ is sometimes referred to as the information measure of $S$ \citep{grunwald2004game}. For us, it measures the ability of a set of data to predict other elements in that particular set of data, or self-similarity. 

We now discuss a few options for scoring rules. The CRPS is given by
\begin{align*}
    S^{\text{CRPS}}(F,y)&=\int_{-\infty}^\infty \left( F(z)- \mbI(y\leq z)\right)^2 \mathrm{d} z=\mbE_{z\sim F}|z-y|-\frac{1}{2}\mbE_{z\sim F, z'\sim F}|z-z'|.
\end{align*}
This is a strictly proper scoring rule and the one used in Section \ref{sec:illus}. Sometimes probabilistic predictions are summarized with their mean and variance. Scores that are only based on the mean and variance then can be used to evaluate the goodness of the predictions. The Dawid-Sebastiani score \citep{dawid1999coherent} (DSS) is one example, given by 
\[S^{\text{DSS}}(F,y) = \frac{(\mu_F-y)^2}{\sigma_F^2} + \ln(\sigma_F^2),\]
where $\mu_F$ is the expected value corresponding to $F$ and $\sigma_F^2$ is the variance corresponding to $F$. This has evident connections to the log-likelihood of a normal distribution, i.e.,
$\ln\left(\exp\left(\frac{-(\mu_F-y)^2}{2\sigma_F^2}\right)/\sqrt{2\pi\sigma^2_F}\right)$, but normality is not required to employ this scoring rule. DSS is a proper scoring rule, but it is not a strictly proper scoring rule. This score can be used, for example, if the only important aspects of the distribution can be distilled down to the mean and variance. A further reduction would simply be the SSE scoring rule:
\[S^{\text{SSE}}(F,y) = (\mu_F-y)^2,\]
(standard trees) which ignores the variance and is especially limiting when the variance is heterogeneous across different subregions of data. 
Lastly, we consider two scoring rules related to two-sided and one-sided intervals. Suppose that we are interested in $(1-\alpha)\times 100\%$ prediction intervals for some $0 \leq \alpha \leq 1$. The two-sided $1-\alpha$ interval score is defined as
\[S^{\text{IS2}}(F,y) = q_F\left(1-\frac{\alpha}{2}\right) -q_F\left(\frac{\alpha}{2}\right) + \begin{cases}
\frac{2}{\alpha} \left(q_F\left(\frac{\alpha}{2}\right) -y\right) & \text{ if } y < q_F\left(\frac{\alpha}{2}\right)\\
 \frac{2}{\alpha} \left(y-q_F\left(1- \frac{\alpha}{2}\right)\right) & \text{ if } y > q_F\left(1-\frac{\alpha}{2}\right) \\
0 & \text{ otherwise, }
\end{cases}\] 
where $q_F(1-\alpha) = \inf \left\{z \in \mbR: 1-\alpha \leq F(z) \right\}$ is the $(1-\alpha)$-th quantile of $F$.
The definition of the quantile is important to maintaining the propriety of the scoring rule. While two-sided intervals are reported for many estimates, risk analysis often focuses on a single upper bound. One-sided intervals are also useful for positive data when the lower bound for a two-sided interval is close to zero. An upper bound interval score (IS1) has the form
\[S^{\text{IS1}}(F,y) = q_F(1-\alpha) + \begin{cases}
 \frac{1}{\alpha} (y-q_F(1-\alpha)), & \text{ if } y > q_F(1-\alpha)\\
0 & \text{ otherwise.}
 \end{cases}\]
A use case of IS1 is when forecasting potential crop yield where we want to find upper bounds to locate high-risk areas. This list of scoring rules is purposefully not exhaustive but presents various circumstances where each can be used. We will use each of these scoring rules to illustrate ideas throughout this article. 

\section{Building a Tree via Scoring Rules}\label{sec:algorithm}
We now formalize the proposed methodology to build a prediction tree based on data consisting of $p$ predictors and a response for each of $n$ observations.


Trees are typically built recursively \citep{breiman1984classification}. Thus the process used to find the first split, i.e., node $t=0$, is mirrored for all subsequent splits. \revision{We let $\mcJ_t$ be the set of indexes of data points that lie in node $t$ (i.e., satisfy the union of splitting rules of node $t$'s parent and grandparents recursively until reaching the root note). A split $(s,k)$ creates two index sets, namely $\mcL_t(k,s) = \left\{i\ \in \mcJ_t: x_i^k\leq s \right\} \text{ and } \mcR_{t}(k,s) =\left\{i \in \mcJ_t : x_{i}^k > s\right\}$. We propose to choose $(k,s)$ by evaluating the predictive distributions resulting from the split via a scoring rule of interest, i.e., the total score similar to \eqref{eq:total-score}, which can be rewritten as 
\begin{align}
    C_t(k,s)   &=|\mcL_t(k,s)|\ES\left(\Fhat_{\mcL_t(k,s)},\Fhat_{\mcL_t(k,s)}\right)+|\mcR_t(k,s)|\ES\left(\Fhat_{\mcR_t(k,s)},\Fhat_{\mcR_t(k,s)}\right).\label{eq:score-calc}
\end{align} Our splitting rule is selecting a predictor $k$ and split value $s$ that minimize $C_t(k,s)$ for a chosen scoring rule; we denote this rule for node $t$ by $(k_t,s_t)$. }

There is an important property of scoring rules that makes our splitting criteria particularly attractive over alternatives. A tree recursively grown with SSE has a key feature of \emph{monotonicity}. \revision{This means, as computed in \eqref{eq:sse-split}, the SSE is non-increasing after splitting:   
\[\sum_{i \in \mcL_{t}(k,s)} (y_i - \ybar_{\mcL_{t}(k,s)} )^2 + \sum_{i \in \mcR_{t}(k,s)} (y_i-\ybar_{\mcR_{t}(k,s)} )^2 \leq \sum_{i\in\mcJ_t} (y_i-\ybar_{\mcJ_t})^2\text{ for all }k\in[p],s\in\mcS^k_t,\]
where $\mcS^k_t$ is the set of all values that the $k$-th predictor takes while being in node $t$. }
Arbitrary splitting rules will not always have this monotonicity property. However, Theorem~\ref{thm:monotone} proves that our proposed splitting criteria have a monotonic feature analogous to SSE. 
\revision{\begin{thm}
Let $\mcJ_t$ contain a subset of indexes in the $t$-th node of the tree. If $S$ is a proper scoring rule, then \[\sum_{i \in \mcL_{t}(k,s)} S\left(\Fhat_{\mcL_{t}(k,s)},y_i\right) + \sum_{i \in \mcR_{t}(k,s)} S\left(\Fhat_{\mcR_{t}(k,s)},y_i\right) \leq \sum_{i \in\mcJ_t} S\left(\Fhat_{\mcJ_t},y_i\right)\text{ for all possible }(k,s).\] That is, any splitting of the data will either reduce the total score or keep it unchanged.\label{thm:monotone}
\end{thm}
 We note, Theorem~\ref{thm:monotone} states that the tree improves prediction on the training data after every split. See Supplemental Material~\ref{suppProofs} for the proof. 
 For a recursive algorithm, such a guarantee to improve the objective by considering more splits prevents the algorithm from getting stuck without finding the best possible tree. }

\subsection{Score-based Trees}
The regression tree via scoring rules, as listed in Algorithm \ref{alg:treebuilding} is constructed starting at the root node with $t=0$,  containing the whole data. At each level $d$ of the tree, all the nodes in that level that were labeled \emph{terminal} are considered to be further split using the splitting criteria \revision{$C_t(k,s)$}, unless they contain fewer than  $N$ (pre-specified parameter) data points, at which point those nodes are labeled as \emph{leaves} and excluded from having offsprings. Ultimately, the leaves will provide the probabilistic predictions for data points that \revision{satisfy the same recursive criteria } that form them. This process repeats up to a pre-specified depth of $D$ in the tree. \revision{$N$ and $D$ are hyperparameters that classically control the tree-based models' complexity. } Each node $t$ that is split will generate two new nodes $2t+1$ and $2t+2$ \revision{with index sets
$\mcJ_{2t+1}:=\mcL_{t}(k_t,s_t)$ and $\mcJ_{2t+2}:=\mcR_{t}(k_t,s_t)$. }

\begin{algorithm}[H]
\small
\caption{{PredictiveTree}{ ($\{( x_i,y_i)\}_{i=1}^n$, max tree depth $D$, min node size $N$})} \label{alg:treebuilding}
\begin{algorithmic}[1]
\State Create a terminal node with indexes in $\mcJ$ containing all data and set depth $d=1$.
    \While{$d\revision{\leq}D$} 
        \For{nodes $t\in\{2^{d}-1,\cdots,2^{d+1}-2\}$ labeled terminal}
            \If{terminal node has at most $N$ data points}
	            \State Label node as leaf and go to next terminal node.
            \Else
                \revision{\State Find $(k_t,s_t)=\argmin_{k\in[p],s\in\mcS_t^k}C_t(k,s)$, where $C_t(k,s)$ is defined in~\eqref{eq:score-calc}.
            \State Create two terminal nodes whose sets of indexes are $\mcL_{t}(k_t,s_t)$ and $\mcR_{t}(k_t,s_t)$. }
            \State Index the two new nodes $2t+1$ and $2t+2$ and set $t=t+1$.
        	    \EndIf
        \EndFor
        \State Set $d = d + 1 $.
    \EndWhile
\end{algorithmic}
\end{algorithm}

\subsection{Parameters and Implementation Specifics}\label{sec:parameters}
Through standard mechanisms \citep[pg 308]{hastie2009elements} in trees, the maximum depth $D$ parameter implies terminal nodes will not be split when they have a certain number of parents. When data is abundant, deeper trees could make the defining halfspaces in the leaves more complicated and in some sense, following probabilistic patterns too closely in the training set at the risk of overfitting. $D$ best scales logarithmically with $n$~\citep{klusowski2020sparse}, which can be tuned with pruning as we will describe later. 

Besides the choice of $D$, because our tree will use ECDFs of $y$ as predictive distributions, it is important to ensure that the minimum number of members of a terminal node is bigger than some $N$. If $N$ is too small, the ECDFs will be poor predictive distributions, especially in the distribution's tails. One rule of thumb for $N$ is the Dvoretzky-Kiefer-Wolfowitz inequality \citep{dvoretzky1956asymptotic,massart1990tight}. This inequality can be inverted to find that $N \geq \frac{\log(2)-\log(\alpha)}{2\varepsilon^2}$ can guarantee at least $\varepsilon$-accurate ECDF with $1-\alpha$ confidence. For example, $95 \%$ confidence at an accuracy of $10 \%$ gives at least 66 samples. In SSE-based trees, however, $N$ is often chosen to be smaller ($\sim 10$~\citep{bertsimas2019optimal}). This can be explained by non-SSE-based trees tending to successfully assess the distributional behavior of the data at the cost of forcing larger terminal nodes. But larger terminal nodes mean smaller trees, which may be advantageous for generalization~\citep{athey2019generalized}. \revision{Importantly, $N$ is not a termination criterion for the tree; it prevents a certain branch of the tree from growing. In all classical tree building literature, both $N$ and $D$ are used to mitigate risks of overfitting. If the tree is too deep, it will tightly track the training data. On the other hand, if a node is too small, it yields too crude ECDF and error-prone statistical information. Controlling the node size with maximum depth $D$ is not guaranteed because while deeper trees ultimately result in smaller nodes, it is still likely that the tree finds small nodes at the earlier depths. Hence, ensuring at least $N$ data points in leaves becomes necessary.}

Given that this algorithm is likely to be used on tall datasets with potentially sizeable \revision{$\mcS_{t}^k$ sets in a terminal node $k$, cycling through all unique values of $\mcS_{t}^k$ (to consider them as a potential split value) leads to a slowdown in the algorithm. Thus for each predictor, one can opt for a search through a set of $1/\ell$ quantiles $\mcQ_{t}^k(\ell)=\{q_t^k(\ell),q_t^k(2\ell),\ldots,q_t^k(1-\ell)\}$ of each predictor $k$ in node $t$ instead. }For example, when $\ell=0.05$, then for each predictor only 20 split values will become candidates to identify the split. 
For discrete predictors with $10$ unique values or less, as well as the categorical predictors, all the possibilities will be considered in the search for best splits. 
\revision{In the experiments, DSS and IS1 have computational time comparable with SSE but CRPS is computationally more expensive. As the last practical consideration, given that CRPS requires $\mcO(n^2)$ operations in~\eqref{eq:crps_diff} and expensive for larger datasets, it is more appropriate for implementation of CRPS-based trees to use an alternative computation of CRPS with $\mcO(n\log n)$ complexity with the approximation $S^\text{CRPS}(\Fhat_{\mcJ},y)\approx\frac{2}{n^2}\sum_{i=1}^{n}(y_{(i)}-y)(n\mbI(y<y_{(i)})-i+\frac{1}{2})$ that uses the order statistics $y_{(i)}$'s (sorted samples) for computation~\citep{zamo2018estimation}. }

\subsection{Pruning Probabilistic Trees} 
\label{sec:pruning}

The tree in Algorithm~\ref{alg:treebuilding} is grown to depth $D$ symmetrically. However, given the greediness of optimal splits, the best tree structure that divides the data into partitions may not be symmetric depending on the identified first optimal split. Trees tend to overfit, and the tree size (i.e., the number of terminal nodes in the tree with depth $D$) is controlled by a complexity (regularization) parameter $\kappa$. Smaller trees are understood to provide better accuracy and interpretability power. Pruning is done after growing a full tree (post-pruning) or simultaneously (pre-pruning), which implies stopping the growth at a node. Pre-pruning is more cost-effective, and its common approaches are listed in \revision{the Supplementary Material Section~\ref{suppA} } for the reader's reference. 

Unlike the common approach, which is growing the tree to its full size and then cutting back subtrees to combine some of the predictions, we explore stopping the tree growth at the nodes whose split does not dramatically improve the prediction quality. There have been setbacks about this approach for potentially missing a very good split that follows a seemingly weak split in the tree \citep{james2013introduction}. However, we adopt this pruning approach to avoid unnecessary computation and obtain smaller trees, albeit with varying sensitivity levels across different scoring rules, which we will explore.

For each terminal node $t$ with more than $N$ data points, \revision{ the optimal split leads to two new terminal nodes that by the monotonicity property satisfy  
\[|\mcJ_{2t+1}|\ES(\Fhat_{\mcJ_{2t+1}},\Fhat_{\mcJ_{2t+1}})+|\mcJ_{2t+2}|\ES(\Fhat_{\mcJ_{2t+2}},\Fhat_{\mcJ_{2t+2}})\leq |\mcJ_{t}|\ES(\Fhat_{\mcJ_{t}},\Fhat_{\mcJ_{t}}).\] Let $\Delta_t:=|\mcJ_{t}|\ES(\Fhat_{\mcJ_{t}},\Fhat_{\mcJ_{t}})-\left(|\mcJ_{2t+1}|\ES(\Fhat_{\mcJ_{2t+1}},\Fhat_{\mcJ_{2t+1}})+|\mcJ_{2t+2}|\ES(\Fhat_{\mcJ_{2t+2}},\Fhat_{\mcJ_{2t+2}})\right)$ be the reduction is score after splitting in node $t$.



By expecting that $\Delta_t$ gradually decreases as the tree becomes deeper, } we propose a heuristic to accept the split on node $t$ if the point-average reduction in the score as a result of it is at least $\kappa\in[0,1]$ factor of the point-average reduction in the score as a result of the split in the root node (the first optimal split), i.e., $\Delta_t/n_t >\kappa \Delta_0/n$ where $n_t=|\mcJ_t|$. Equivalently, we accept the best split at node $t$ if 
\begin{align}
    \ES(\Fhat_{\mcJ_{t}},\Fhat_{\mcJ_{t}})-&\left(\frac{n_{2t+1}}{n_t}\ES(\Fhat_{\mcJ_{2t+1}},\Fhat_{\mcJ_{2t+1}})+\frac{n_{2t+2}}{n_t}\ES(\Fhat_{\mcJ_{2t+2}},\Fhat_{\mcJ_{2t+2}})\right)\nonumber \\
    &>\kappa\left(\ES(\Fhat_{\mcJ},\Fhat_{\mcJ})-\left(\frac{n_{1}}{n}\ES(\Fhat_{\mcJ_{1}},\Fhat_{\mcJ_{1}})+\frac{n_{2}}{n}\ES(\Fhat_{\mcJ_{2}},\Fhat_{\mcJ_{2}})\right)\right).\label{eq:pruning} 
\end{align} 
Note, with $\kappa = 0$, Algorithm~\ref{alg:treebuilding} remains the same. As $\kappa$ increases, the size of the tree becomes smaller. If $\kappa=1$, we only have a root node in the tree.

\revision{\section{Near-Optimality of the Empirical Split}\label{sec:empirical-split}
This section explains some of the theoretical behavior of our trees learned from finite data. Our treatment will be decidedly less general than comparative work on the asymptotic behavior of trees \citep{gordon1980consistent,toth2011building,scornet2015consistency}. This section's goal is to explain the impact of finite data on the new splitting criteria based on scoring rules. With some loss of generality, this section will only consider the behavior of a single split and keeps the available dataset used for splitting fixed (not random). Here we answer the following question in a general setting: given that our split is based on finite data, how does this compare to the prediction if one chooses the split optimally? 

Say that we have a collection of realizations $(x_1,y_1),\ldots,(x_n,y_n)$ which are assumed to be from some joint distribution. Throughout the analysis, we fix this dataset that has an optimal split (yielding lowest total score when used to predict unseen targets $y$). Denote the potential splits by regions $A_1,\ldots,A_t,\ldots, A_T$; these are a collection of \emph{half-spaces} of the form $\{x:\ x^k\leq s\}$. The potential splits are considered to be nonrandom for simplicity.  In this section, we replace $(k,s)$ splits with $A$ regions to ease the exposure, and use $\mcL(A;n)$ and $\mcR(A;n)$ to reflect the dependence on $n$. Our chosen split is dictated by 
\[\hat{A}_n = \argmin_{A \in \{A_1,\ldots, A_T\}} \sum_{x_{i} \in A } S\left(\widehat{F}_{\mcL(A;n)} ,y_i\right) + \sum_{x_{i} \in A^c } S\left(\widehat{F}_{\mcR(A;n)},y_i\right), \] 
where $\mcL(A;n)$ and $\mcR(A;n)$ are the subsets of $n$ data points with their predictors lying on either side of the split that defines sub-region $A$. Let $F_{\mcL(A;\infty)}$ and $F_{\mcR(A;\infty)}$ represent the \emph{true} conditional distributions of $y$ for data whose predictors lie on either side of the split that defines sub-region $A$. It makes sense to judge a split $A$ via the following criteria
\[g (A) := \text{ES}\left(\widehat{F}_{\mcL(A;n)},F_{\mcL(A;\infty)}\right) \mbP \left(x \in A\right)+ \text{ES} \left(\widehat{F}_{\mcR(A;n)},F_{\mcR(A;\infty)} \right) \mbP \left(x \in A^c\right). \]
This represents the expected score for a new prediction of unobserved data after the split is finished. 
An oracle would choose the split such that
\[g^* := g(A^*_n) = \min_{A \in \{A_1,\ldots, A_T\}} g(A), \] where the \emph{oracle split} choice that yields $g^*$ is denoted by $A_n^*$. Clearly, we would like $g\left(\hat{A}_n\right)$ to be as close as possible to $g^*$. 
In the spirit of the generality of this article, we now state a condition for general scoring rules. 
\begin{thm} \label{thm:consist_gen}
Let $y_1,\ldots,y_{n}$ be independent draws from a mixture of two distributions $F$ and $G$. Let $n_F$ be the number drawn from $F$ and $n_G$ be the number drawn from $G$. Let $\widehat{F}_n$ and $\widehat{G}_n$ represent our empirical predictive distribution based on the points drawn from each distribution. Let $\mathcal{P}$ be a class of all distributions that includes every distribution of a random $y$ given $x \in A$ for all subsets $A$ of the predictor space. If for all $F,G \in \mathcal{P}$, 
\begin{equation}
    \text{ES} \left(\frac{n_F}{n_F + n_G} \widehat{F}_n +\frac{n_G}{n_F + n_G} \widehat{G}_n ,\widehat{F}_n\right) \inP \text{ES} \left(\frac{n_F}{n_F + n_G} \widehat{F}_n +\frac{n_G}{n_F + n_G} \widehat{G}_n,F\right),
    \label{eq:convergence-condition}
\end{equation}
 as $n \rightarrow \infty$, then $g(\hat{A}_n) \inP g^*$ as $n \rightarrow \infty$.
\end{thm}
Theorem~\ref{thm:consist_gen} states that the predictive distributions (ECDFs in subregions, given a fixed dataset) of a score-based tree approach the highest accuracy (smallest score) when predicting increasingly large sets of unseen data. The implication of~\eqref{eq:convergence-condition} is that the score must obey consistency (in the second argument) for the target variable. For the special cases of scoring rules used in this paper, the next corollary shows this requirement is met in some reasonably well-behaved probability space $\mathcal{P}$. The tricky part of showing this result for a given scoring rule is that $\widehat{F}_n$ appears on both sides of the score. Thus we cannot directly invoke the law of large numbers. See Supplemental Materials~\ref{suppProofs} for the proofs.



Besides providing the result in full generality, we next offer specific conditions for the scoring rules introduced in Section~\ref{sec:background_score}. 
\begin{cor} \label{thm:consist_CRPS}
If $S$ is chosen to be CRPS or DSS, then assuming that for all subregions of predictor space $A$, the distribution of a random $y$ conditioned on $x \in A$ is such that $\mathbb{E} (y^2)$ is finite, we get $g(\hat{A}_n) \inP g^*$ as $n \rightarrow \infty$. If $S$ is chosen to be IS1 or IS2,  then assuming that for all subsets of predictor space, $A$, the distribution of a random $y$ conditioned on $x \in A$ is such that the CDF for $y$ is strictly increasing near $\alpha/2$ and $1-\alpha/2$ for IS2 and $1-\alpha$ for IS1, we get $g(\hat{A}_n) \inP g^*$ as $n \rightarrow \infty$.
\end{cor}
The moment condition of Corollary \ref{thm:consist_CRPS} gives guarantees that the CRPS/DSS score is well-behaved. For the interval and upper bound score, the condition shifts from a moment-based condition to one that guarantees convergence of the sample quantile. This condition can be modified for discrete data.} 

These results are intended to verify the intuition that these scores based on ECDFs lead to splits that, even though we have no proof for them to be the correct optimal splits, their resulting tree scores will be close enough to the scores in the optimal trees with high probability. Thus, the scoring rule choice will impact the ultimate tree that is constructed, no matter how much data is present. The choice of scoring rule thus cannot be ignored and can have a large impact on the resulting prediction. One example of this was in Section \ref{sec:illus}, but our analysis of real data in Section \ref{sec:numerics} confirms this result. \revision{Table~\ref{tab:split-interpret} in Supplemental Material Section~\ref{suppH} also shows, using synthetic datasets, that certain scoring rules fall short of finding the boundaries in the data where the probabilistic behavior changes especially if the change happens less obviously and beyond mean values.}


\revision{\section{Numerical Experiments}}\label{sec:numerics}
\revision{In this section, we examine the new tree construction methods using different scoring rules with experiments on synthetic datasets and real public datasets. As a baseline for comparison, we use standard trees with SSE criteria. All approaches are implemented under our own Python package \texttt{scoreTree}, publicly available at \url{https://github.com/sshashaa/scoreTree}. The code is also provided as an online supplementary material and the \texttt{README} file provides instructions to replicate examples from the paper.}

\subsection{Synthetic Datasets}
\begin{table}[htp]\caption{Synthetic datasets with \texttt{logNormal(lgN)} distributions of $y$ on $x$ subregions.}\label{tab:synth-data}
\centering
\footnotesize
\setlength{\tabcolsep}{1pt}
\begin{tabular}{c | l | c | c | c | c | c} 
\cline{2-7}
& Regions & $-1 < x < -0.5$& $-0.5 < x < 0$ & $0 < x < 0.5$ & $0.5 < x < 1$ & Sub-region boxplots\\ \hline	
\multirow{4}{*}{\rotatebox[origin=c]{90}{\bf Easy Dataset}} & $y$ Dist. & \texttt{lgN(2,1/2)}& \texttt{lgN(3,1/3)} & \texttt{lgN(4,1/4)} & \texttt{lgN(5,1/5)} & \multirow{4}{*}{
\resizebox{!}{2.4cm}
 {\includegraphics[width=0.4\linewidth]{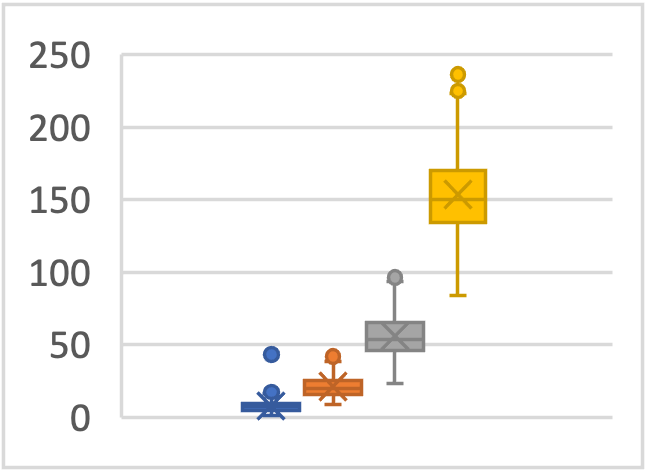}}
 } \\
\cline{2-6}
& $\mbE[y]$ & 7.9 & 21.1 & 56.1 & 153.4 & \\
\cline{2-6}
& $\mbE[y^2]$ & 82.4 & 494.0 & 3359.4 & 24372.0 &\\
\cline{2-6}
& $\mbE[y^3]$ & 1210.3 & 12894.8 & 213981.9 & 4012557.9 &\\
\hline\hline
\multirow{4}{*}{\rotatebox[origin=c]{90}{\bf Hard Dataset}} & $y$ Dist. & \texttt{lgN(1/2,0.5)}& \texttt{lgN(1/3,0.6)} & \texttt{lgN(1/4,0.3)} & \texttt{lgN(1/5,0.3)} & \multirow{4}{*}{
\resizebox{!}{2.4cm}
 {\includegraphics[width=0.4\linewidth]{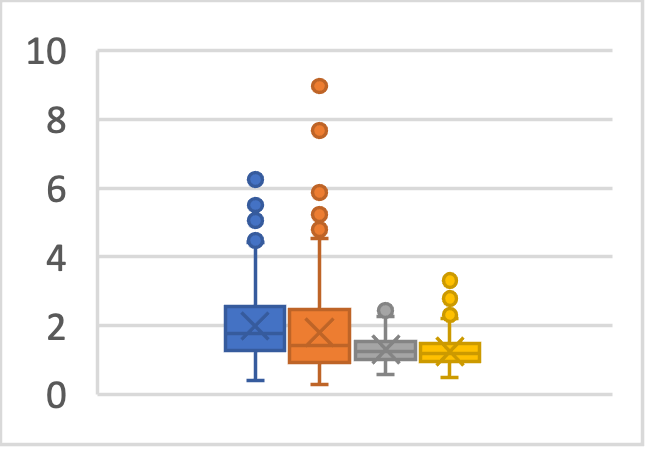}}
 }
 \\\cline{2-6}
& $\mbE[y]$ & 1.99 & 1.80 & 1.31 & 1.26 &
 \\\cline{2-6}
& $\mbE[y^2]$ & 5.04 & 4.73 & 1.88 & 1.76 & \\\cline{2-6}
& $\mbE[y^3]$ & 15.66 & 17.50 & 2.90 & 2.75 &\\
\hline
  \end{tabular}
\end{table}

Two synthetic datasets for one-dimensional continuous feature space in $[0,1]$ are designed with response behavior in four regions described in Table~\ref{tab:synth-data}. The \emph{easy} dataset exhibits easy-to-distinguish behavior of the response in each subregion, evidenced by significant differences in the first and higher central moments. The conjecture is that SSE should easily separate these regions using the first moment. On the other hand, the \emph{hard} dataset entails more similarly behaving responses in the first two moments everywhere, making it harder for SSE-based trees to predict when there is difference in behavior. \revision{Although real-world data may not be in a tree structure, the synthetic datasets mimic the heteroscedasticity and responses that follow a mixture of distributions.}

\begin{algorithm}[H]
\small
\caption{ScoreTreeExperiment(bootstrapped datasets $\ell_b,\ b=1,2,\ldots,r$)} \label{alg:experiments}
\begin{algorithmic}[1]
\For{Bootstrap $\ell_b,\ b=1,2,\ldots,r$}
    \For{Pruning parameter $\kappa\in\{0,0.1,0.3,0.5,0.8\}$}
        \For{Scoring rule $\text{Build} \in \{\text{SSE, CRPS, DSS, IS1}\}$}
            \State Train a tree with the $\text{Build}$ score, $\ell_b$ data, and pruning parameter $\kappa$. 
            \State Return $t(j;\text{Build},\kappa,b)$, terminal node containing $j$-th data point $\forall j\in\ell_b$.
                \For{Scoring rule $\text{Eval} \in \{\text{SSE, CRPS, DSS, IS1}\}$}
                    \State Compute $\text{I}_{b}^{\text{Eval}}(\text{Build},\kappa):=\sum_{j\in\ell_b}S^{\text{Eval}}\left(\Fhat_{\mcJ_{t(j;\text{Build},\kappa,b)}}, y_j\right)$.
                    \State Compute $\text{O}_{b}^{\text{Eval}}(\text{Build},\kappa):=\sum_{j\notin\ell_b}S^{\text{Eval}}\left(\Fhat_{\mcJ_{t(j;\text{Build},\kappa,b)}},y_j\right)$.
                \EndFor
       \EndFor
    \EndFor
\EndFor
                    
\end{algorithmic}
\end{algorithm}
We construct trees with several scoring rules (Build$ \in\{\text{SSE, CRPS, DSS, IS1}\}$) and evaluate their performance under a varied number of observations $n$ and different choices of the pruning parameter $\kappa$ introduced in (\ref{eq:pruning}). \revision{The benchmark procedure is summarized in Algorithm~\ref{alg:experiments}. For each dataset presented in Table~\ref{tab:synth-data}, samples of size $n\in\{200,400,800,1600\}$ are generated as training data sets and thresholds $\kappa\in\{0.0,0.1,0.3,$ $0.5,0.8\}$ are implemented with each tree. For the comparisons, for each experiment (i.e., for each combination of Build, $n$, and $\kappa$), we generated one test set of 1,000 observations (to evaluate its performance) and 30 training data sets of size $n$ (to build the tree), the latter representing 30 replicates ($r = 30$) of the experiment. On each replicate, we fit trees with Build $ \in\{\text{SSE, CRPS, DSS, IS1}\}$ score to the training data (see Line~3). Each tree is evaluated with both in-sample (I) and out-of-sample (O) errors via different scores denoted by Eval $ \in\{\text{SSE, CRPS, DSS, IS1}\}$ using training and test sets, respectively.  Since the responses are non-negative, we only use the upper interval score $\text{IS1}$ with $\alpha=0.2$ (implying that, when fitting a tree we penalize a prediction that is worse than the 0.8-quantile of the predictive distribution). For all trees, $D = 4$, $N = 50$, $\ell = 0.05$ following the rules of thumb described in Section~\ref{sec:parameters}. Data is repeatedly ($r=30$ independent times) divided into an equal-sized training set for all experiments with common random numbers (CRN). CRN helps us see the effect of different trees and their performances on the same sets of data for training and testing, reducing the variability for comparison. } Consequently, the predictive distributions are approximated by the data points that lie in the terminal node $t(\cdot;\cdot)$ as indicated in Line~5 of Algorithm~\ref{alg:experiments}. Across $r=30$ replicates, in-sample and out-of-sample errors, represented by $\{\text{I}_{b}^{\text{Eval}}(\text{Build},\kappa)\}_{b=1,2,\cdots,r}$ and $\{\text{O}_{b}^{\text{Eval}}(\text{Build},\kappa)\}_{b=1,2,\cdots,r}$, \revision{for the trees constructed with $\text{Build}$ score and pruned with threshold $\kappa$ are evaluated with Eval score to summarize the results (see Lines 7--8 in Algorithm~\ref{alg:experiments}). }

Our first comparison validates whether the tree built with a scoring rule of interest (Eval) yields better probabilistic predictions (lower scores) on out-of-sample data than trees constructed with the same data but with different scoring rules. \revision{We evaluate the paired difference of scores for out-of-sample scores:
\begin{equation}
    \text{OD}_{b}(\text{Eval},\text{Build}, \kappa):=\text{O}_{b}^{\text{Eval}}(\text{Eval}, \kappa)-\text{O}_{b}^{\text{Eval}}(\text{Build}, \kappa),\ \forall b=1,2,\cdots,r\label{eq:paired-diff}
\end{equation} between trees constructed with $\text{Eval}$ and $\text{Build}$ scores using the Eval score, 
where negative values validate that trees yield better predictions if trained with the same scoring rule that evaluates them (based on the goal of prediction).

\begin{figure}[tb]
  \centering
  \includegraphics[width=.6\linewidth]{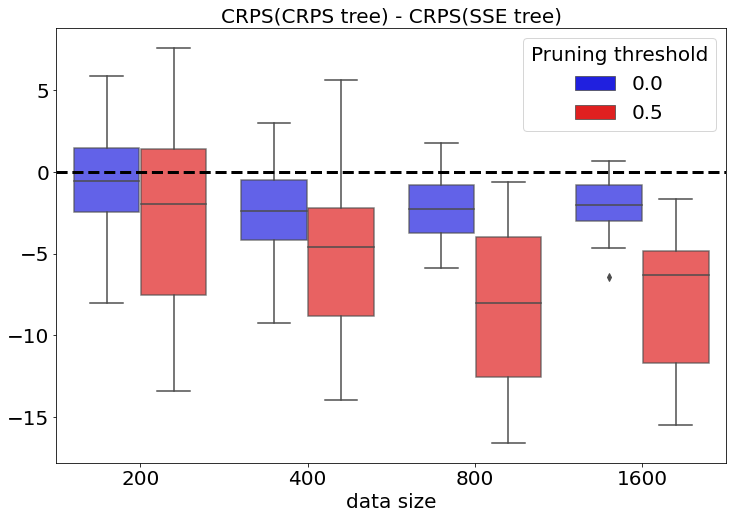}
\caption{Boxplots of the paired difference of CRPS scores between CRPS trees and SSE trees on out-of-sample predictions for the hard dataset suggest that with the growing size of training data, CRPS trees provide better predictions than SSE trees. 
}
\label{fig:crps-hard}
\end{figure}

Figure~\ref{fig:crps-hard} shows one instance of these comparisons with Build = SSE and Eval = CRPS for the hard dataset with two choice for pruning, $\kappa \in \{0, 0.5\}$. 
We observe that as the training data size increases, SSE-based trees fail to provide good predictions (when the goal is to have a good CRPS performance). This weakness of SSE-based trees is statistically significant with pruning. See Table~\ref{tab:hypotest} in the Supplemental Material for a complete statistical test for all pairs of Eval and Build scores. This complete statistical test suggests that we can generally validate that Eval trees are better than Build trees when compared in Eval score. However, for the hard dataset, some scores struggle more than others. An interesting observation is the effect of pruning in helping the fit when using different scores on both datasets. For example, for the hard dataset, we observe that even a small pruning of $\kappa= 0 .1$ can impact the validation of DSS- and IS1 trees.}

\begin{figure}[H]
	\centering
	\footnotesize
	\begin{forest}
    label0/.style={
    tikz+={
      \node [anchor=mid east] at ([xshift=-3.8cm].west |- .mid) {#1};
    },
  },
    label1/.style={
    tikz+={
      \node [anchor=mid east] at ([xshift=-0.5cm].west |- .mid) {#1};
    },
  },
  [\text{}, s sep = 5mm,  rectangle, draw
		[\text{}, s sep=2mm, l = 15mm, rectangle, draw, edge label={node[midway, left, font=\scriptsize]{$< -0.02$}}
			[\text{}, s sep=6mm, l = 20mm, rectangle, draw, edge label={node[midway, left, font=\scriptsize]{$ < -0.50$}}
				[\text{}, s sep=9mm, l = 20mm, edge=dashed, rectangle, draw, edge label={node[midway, left, sloped, above, font=\scriptsize]{$<-0.95$}}
					[\text{}, s sep=8mm, l = 20mm, rectangle, draw, edge=dashed, edge label={node[midway, left, sloped, above, font=\scriptsize]{$<-0.97$}}]
					[\text{}, s sep=8mm, l = 20mm, rectangle, draw, edge=dashed, edge label={node[midway, left, sloped, above, font=\scriptsize]{$\geq-0.97$}}]]
				[\text{}, s sep=9mm, l = 20mm, edge=dashed, rectangle, draw, edge label={node[midway, left, sloped, below, font=\scriptsize]{$\geq-0.95$}}
					[\text{}, s sep=8mm, l = 20mm, rectangle, draw, edge=dashed, edge label={node[midway, left, sloped, above, font=\scriptsize]{$<-0.93$}}]
					[\text{}, s sep=8mm, l = 20mm, rectangle, draw, edge=dashed, edge label={node[midway, left, sloped, above, font=\scriptsize]{$\geq-0.93$}}]]]
			[\text{}, s sep=6mm, l = 20mm, rectangle, draw, edge label={node[midway, left, font=\scriptsize]{$x \geq -0.50$}}
				[\text{}, s sep=9mm, l = 20mm, edge=dashed, rectangle, draw, edge label={node[midway, left, sloped, below, font=\scriptsize]{$< -0.38$}}
					[\text{}, s sep=8mm, l = 20mm, rectangle, draw, edge=dashed, edge label={node[midway, left, sloped, above, font=\scriptsize]{$<-0.43$}}]
					[\text{}, s sep=8mm, l = 20mm, rectangle, draw, edge=dashed, edge label={node[midway, left, sloped, above, font=\scriptsize]{$\geq-0.43$}}]]
				[\text{}, s sep=9mm, l = 20mm, edge=dashed, rectangle, draw, edge label={node[midway, left, sloped, below, font=\scriptsize]{$\geq -0.38$}}
					[\text{}, s sep=8mm, l = 20mm, rectangle, draw, edge=dashed, edge label={node[midway, left,  sloped, above, font=\scriptsize]{$<-0.37$}}]
					[\text{}, s sep=8mm, l = 20mm, rectangle, draw, edge=dashed, edge label={node[midway, left,  sloped, above, font=\scriptsize]{$\geq-0.37$}}]]]]
		[\text{}, s sep= 2mm, l = 15mm, rectangle, draw, edge label={node[midway, left, font=\scriptsize]{$x \geq -0.02$}}
			[\text{}, s sep=6mm, l = 20mm, rectangle, draw, edge label={node[midway, left, font=\scriptsize]{$x < 0.51$}}
				[\text{}, s sep=9mm, l = 20mm, edge=dashed, rectangle, draw, edge label={node[midway, left, font=\scriptsize]{$x < 0$}}]
				[\text{}, s sep=9mm, l = 20mm, edge=dashed, rectangle, draw, edge label={node[midway, right, font=\scriptsize]{$x \geq 0$}}
					[\text{}, s sep=8mm, l = 20mm, rectangle, draw, edge=dashed, edge label={node[midway, right, sloped, above, font=\scriptsize]{$< 0.47$}}]
					[\text{}, s sep=8mm, l = 20mm, rectangle, draw, edge=dashed, edge label={node[midway, right, sloped, above, font=\scriptsize]{$\geq 0.47$}}]]]
			[\text{}, rectangle, s sep=6mm, l = 20mm, draw, edge label={node[midway, right, font=\scriptsize]{$x \geq 0.51$}}
				[\text{}, s sep=9mm, l = 20 mm, rectangle, draw, edge=dashed, edge label={node[midway, right, sloped, above, font=\scriptsize]{$< 0.89$}}
					[\text{}, s sep= 9mm, l = 20 mm, rectangle, draw, edge=dashed, edge label={node[midway, right, sloped, above, font=\scriptsize]{$< 0.52$}}]
					[\text{}, s sep= 9mm, l = 20 mm, rectangle, draw, edge=dashed, edge label={node[midway, right, sloped, above, font=\scriptsize]{$\geq 0.52$}}]]
				[\text{}, s sep=9mm, l = 20 mm, rectangle, draw, edge=dashed, edge label={node[midway, right, sloped, above, font=\scriptsize]{$\geq 0.89$}}
					[\text{}, s sep= 9mm, l = 20 mm, rectangle, draw, edge=dashed, edge label={node[midway, right, sloped, above, font=\scriptsize]{$<0.98$}}]
					[\text{}, s sep= 9mm, l = 20 mm, rectangle, draw, edge=dashed, edge label={node[midway, right, sloped, above, font=\scriptsize]{$\geq 0.98$}}]]]]]
	\end{forest}
	\caption{A CRPS tree at $D=4$ with $\kappa=0$ (dashed lines) and $\kappa=0.5$ (solid lines).}
	\label{fig:sample tree}
\end{figure}
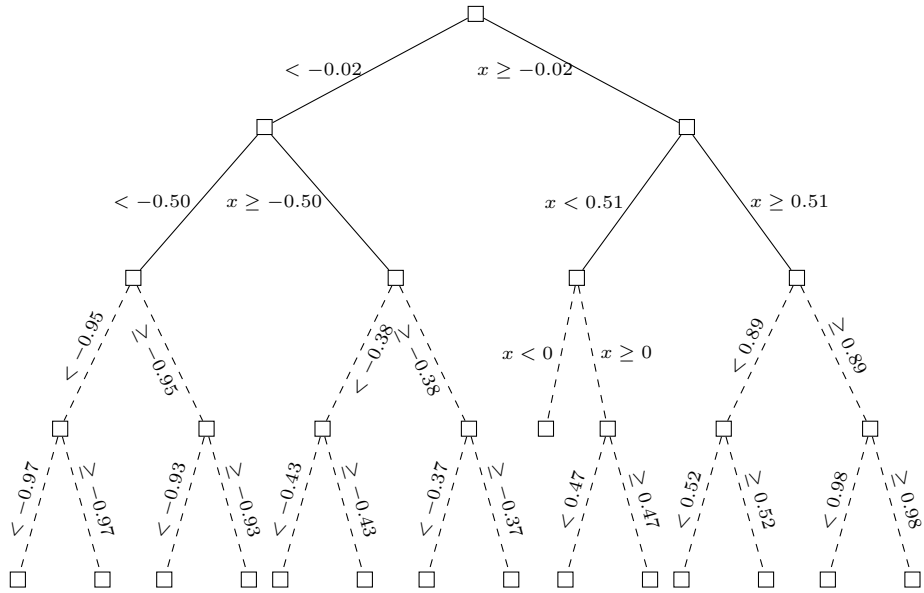
Different scoring rules will best function under varying intensities of pruning. Figure~\ref{fig:sample tree} shows a CRPS tree trained with different pruning parameters for one instance of the hard dataset. As expected, the higher pruning values lead to a smaller tree (solid lines); yet the same pruning parameter may lead to different tree sizes when used with different scores. The best pruning value for SSE may not be the same as that for CRPS. To compare each tree with its counterparts built via other scores, we first find the best pruning value for each score   via cross-validation (using out-of-sample results): $$\kappa^*(\text{Score}):=\argmin_{\kappa}\ \frac{1}{r}\sum_{b=1}^r\text{O}_{b}^{\text{Score}}(\text{Score},\kappa),$$ given a data size. These values are summarized in Table~\ref{tab:kappa-str}. These values suggest that for the easy dataset, $\kappa^*=0.1$ is generally a good value across training data sizes and scoring rules, except SSE which does not appear to benefit from pruning (aligned with evidence from the hypothesis test results in Table~\ref{tab:hypotest} of the Supplementary Material).

\begin{table}[H]
\centering
\footnotesize
\caption{The optimal pruning $\kappa^{*}(\text{Score})$ for each scoring rule and each data size.}\label{tab:kappa-str}
\begin{tabular}{|c|cccc|cccc|}
\cline{2-9}
\multicolumn{1}{c|}{}& \multicolumn{4}{c|}{{\bf Easy Dataset}} & \multicolumn{4}{c|}{{\bf Hard Dataset}}\\ \hline
 \diagbox{ Score}{$n:$ Data size} & 200 & 400 & 800 & 1600 & 200 & 400 & 800 & 1600 \\ \hline
 $\text{SSE}$ & 0.0 & 0.0 & 0.0 & 0.0 & 0.8 & 0.8 & 0.8 & 0.3\\ \hline
 $\text{CRPS}$ & 0.1 & 0.1 & 0.1 & 0.0 & 0.8 & 0.8 & 0.8 & 0.5 \\ \hline $\text{DSS}$ & 0.1 & 0.1 & 0.1 & 0.1 & 0.8 & 0.3 & 0.1 & 0.1 \\ \hline $\text{IS1}$ & 0.1 & 0.1 & 0.1 & 0.0 & 0.8 & 0.8 & 0.8 & 0.3\\ \hline
\end{tabular}
\end{table}

There are more irregularities in the hard dataset. All scoring rules favor pruning, some less than others when sufficient training data is available. However, for the small data size, all scoring rules provide their best performance with the smallest tree that is pruned with $\kappa=0.8$. DSS shows different behavior than the other scores for the hard dataset. Besides these observations, while not visible in Table~\ref{tab:kappa-str}, the variance of the optimal $\kappa$ performance for IS1 is noticeably larger than the other scores. Another noteworthy point is that for the in-sample results, the $\kappa^*=0.0$ for all scores and all data sizes implies that without pruning, the trees are subject to overfit, especially for the hard dataset.

\begin{figure}[H]
\centering
\begin{subfigure}{.25\textwidth}
  \centering
  \includegraphics[width=1\linewidth]{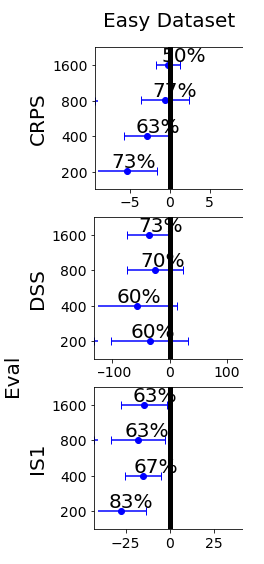}
\end{subfigure}
\begin{subfigure}{.175\textwidth}
  \centering
  \includegraphics[width=1\linewidth]{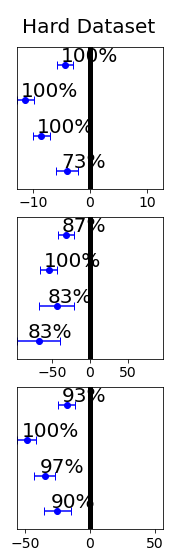}
\end{subfigure}
\caption{Confidence intervals of difference of optimal scores between SSE and other scores and corresponding success probabilities (labels on each interval) for each Eval tree (using the same resample of data for building both trees) across 30 replications with varying training data sizes using the easy and hard datasets.}
\label{fig:summary-plots}
\end{figure}

To alleviate the interactive effect of pruning and scores, we compare the best version of each score-based tree using their corresponding optimal pruning value; typically this pruning value is chosen in a validation step by the user for a given dataset. We construct a confidence interval for the paired difference of optimal scores 
$$\text{OD}_b^*(\text{Eval},\text{Build}):=\text{O}_{b}^{\text{Eval}}(\text{Eval},\kappa^*(\text{Eval}))-\text{O}_{b}^{\text{Eval}}(\text{Build},\kappa^*(\text{Build})),\ \forall b=1,2,\cdots,r$$ defined similar to~\eqref{eq:paired-diff}. 
We also estimate the \emph{probability of success}, defined as the fraction of replications with the Eval tree outperforming the Build tree score in Eval score, i.e.,  $$\widehat\Pr\{\text{Eval succeeds over Build}\}:= \frac{1}{r}\sum_{i=1}^r\mbI\{\text{O}_{b}^{\text{Eval}}(\text{Eval},\kappa^*(\text{Eval})) \leq \text{O}_{b}^{\text{Eval}}(\text{Build},\kappa^*(\text{Build}))\}.$$ Figure~\ref{fig:summary-plots} summarizes $\text{OD}^*(\text{Eval},\text{SSE})$ confidence intervals and success probabilities for easy and hard datasets.



\revision{In most cases, especially for the hard dataset, the out-performance of CRPS-, DSS-, and IS1-based trees over SSE-based trees is statistically significant. The percentage of times that an SSE-based tree is worse than its counterparts is also notably high across cases. This result confirms that non-SSE-based trees can achieve better probabilistic predictions when the data is not completely summarized by mean values (a property synthesized in the hard dataset). We also observe that the length of the confidence intervals often decreases with sample size. This can be explained by the fact that exploiting probabilistic properties (mean, quantiles, variance) is noisier with smaller training data. 

In a follow-up experiment, to see whether there is a score that unanimously outperforms other scores, we investigated  $\text{OD}^*(\text{Eval},\text{Build})$ confidence intervals and success probabilities for building trees with CRPS, DSS, and IS1 scores. Figure~\ref{fig:summary-plots-2} summarizes these results. 
\begin{figure}[H]
\begin{subfigure}{.52\textwidth}
  \centering
  \includegraphics[width=1\linewidth]{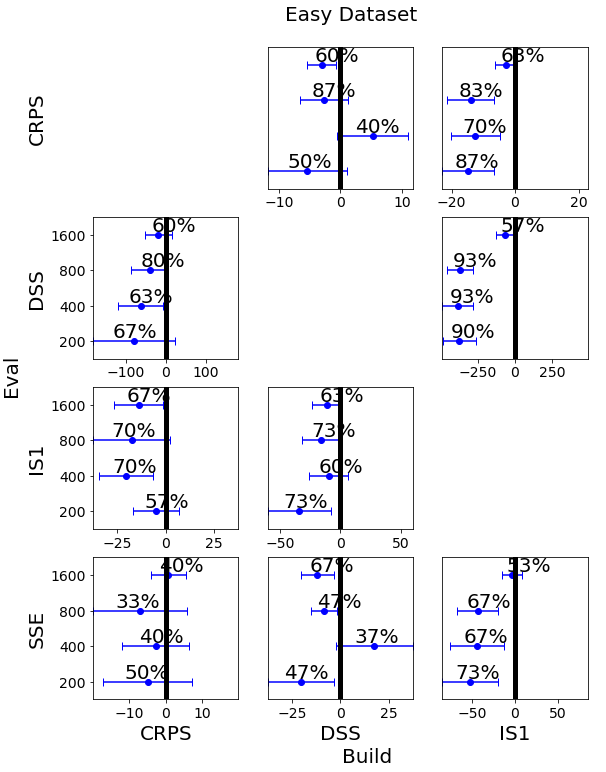}
\end{subfigure}
\begin{subfigure}{.47\textwidth}
  \centering
  \includegraphics[width=1\linewidth]{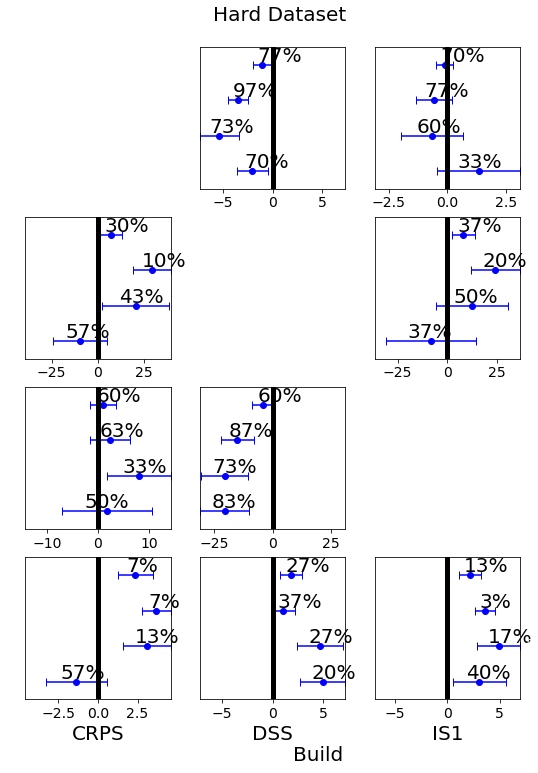}
\end{subfigure}
\caption{Confidence intervals of paired difference of optimal scores and success probabilities for each Eval tree and its Build tree counterpart across 30 replications with varying training data sizes for the easy and hard datasets.}
\label{fig:summary-plots-2}
\end{figure}

A number of observations from Figure~\ref{fig:summary-plots-2} are noteworthy:
\begin{itemize}
    \item[(a)] In the easy dataset, all scoring rules provide relatively similar probabilistic predictions; while IS1 almost never leads to better trees (regardless of the goal of prediction), there is not enough statistical evidence to say the same for CRPS and DSS scores.
    \item[(b)] In the hard dataset, CRPS- and IS1-based trees provide similar performance to one another. But compared to DSS- and SSE-based trees, they are more likely to provide better predictions and their improved performance is statistically significant as the data size increases. This is expected because with more data the empirical estimates of quantiles that are needed in both of these scoring rules become more accurate. DSS-based tree only show better performance compared to SSE-based trees, but do so with statistical significance invariably across data sizes. Non-SSE trees can lead to better SSE scores in almost all cases. The same holds true for DDS score when the training data size is not too small. Good CRPS and IS1 scores are not achievable with SSE- and DSS-based trees. 
\end{itemize}

In our final investigation of this section, we compare the trees' ability to find the correct splits. While the main purpose of score-based trees is to produce better probabilistic predictions, identifying the correct subregions will render their suitability more convincing. As expected the non-SSE trees can be more successful in identifying the subregions; Figure~\ref{fig:sample tree} shows, for example, that the CRPS tree is able to identify correct splits (within a $\pm0.02$ margin of error). If the tree is not sufficiently pruned, many incorrect splits will be contained in the tree structure (dash lines). But even in a sufficiently small tree, the split values can be incorrect if other scores are used for splitting. For a more comprehensive comparison in this regard, see Table~\ref{tab:split-interpret} in Supplementary Material Section~\ref{suppH}. Table~\ref{tab:split-interpret} shows that (i) all true split points are more likely to be recovered by non-SSE scores, (ii) SSE and DSS trees tend to find more incorrect splits in the data, while CRPS trees find the fewest incorrect splits on average, and (iii) among non-SSE scores, the subregions that are more difficult to identify are more likely recovered by IS1 than DSS, and most likely recovered by CRPS. A direct implication of these correct identifications of split values is the improved interpretability of data. In many applications such as in health outcome predictions, these correct split values lead to correct clustering of patients with distributionally similar outcomes and more accurate personalized predictions~\citep{mao2022personalized}. 
}

\subsection{Real Datasets}
We also investigate the score-based trees on two real datasets; see \revision{Supplementary Material Section~\ref{supprealdata} for their descriptions. } The first is the yield data from the Ethiopian Annual Agricultural Surveys with 174,028 rows $\times$ 5 predictors, and $\sim$94K unique response values. The second is the Divvy bikeshare data from the city of Chicago with 1.3M rows $\times$ 9 predictors and $\sim$3.4K unique response values. Our analysis again entails $r=30$ replications, with training data of sizes $n=5,000$ and $n=10,000$, and computed $\kappa^*$ for each score. 

\begin{figure}[H]
    \centering
    \includegraphics[width=.45\linewidth]{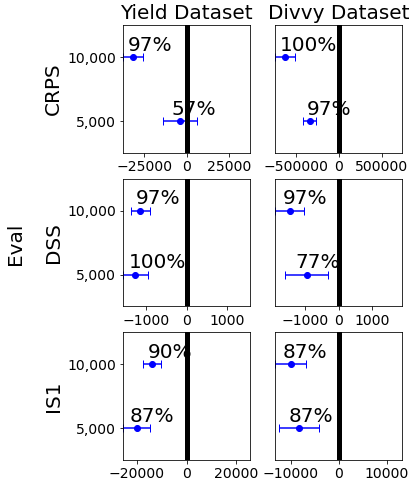}
    \caption{Comparison of SSE trees with other score-based trees for the yield and divvy datasets. The numbers on the lines show the percentage of success across 30 replications.}
\label{fig:yield-data-results}
\end{figure}


We compare the Eval-score of optimal eval trees with those of the optimal SSE trees. Similar to the synthetic data, we make our comparisons with (i) probability of success and (ii) paired difference confidence intervals. Figure~\ref{fig:yield-data-results} illustrates that \revision{non-SSE trees provide statistically better predictions than SSE trees. In most cases, non-SSE trees outperform the SSE trees more evidently with larger training data. }

\section{Concluding Remarks}\label{sec:conclude}
\revision{
In this article, we discuss that standard mechanisms for regression trees are not designed to grow a tree with the goal of creating good nonparametric predictive distributions. We aim to build a tree with generally good predictive distribution and conclude that fitting regression trees to training data by using proper scoring rules other than SSE as the split criteria can improve predictive properties. This is because, unlike SSE that summarizes the predictive distribution with its mean value, other proper scoring rules will focus on various other summary statistics (quantiles, higher moments, etc) that are of importance depending on the application and heterogeneity in the data. 
Since the recursive partitioning of the proposed trees is dictated by the scoring rule, when the scoring rule is chosen to align with the goal of prediction or based on some knowledge about the data, the resulting tree produces improvements over SSE-based trees. The type of score can also affect the additional computation for computing other summary statistics in the predictive distributions, but if chosen well, it can lead to not only better predictions, but potentially also better interpretation on the partitions created for the data (finding the correct split points). 
Our near-optimal analysis and numerical results conclusively show unanimous gain in using scoring-based trees. }
By extension, trees with proper scoring rules can provide a significant improvement when used as based learners and in ensemble settings \revision{such as forests}. We leave these important extensions for future research. 


\bibliographystyle{plainnat}
\bibliography{sara_cit,matt_cit}
\newpage
\input{revised-supp}

\end{document}

%% file: revised-supp.tex
  \title{\LARGE \bf Supplementary Materials}
%
%

\spacingset{1.8} 

The supplemental material includes the supplementary material to the main article ``Building Trees for Probabilistic Prediction via Scoring Rules'', organized in the following order:
    \begin{enumerate}[]
        \item[] Appedix A: List of the existing pre-pruning algorithms.
        \item[] Appedix B: Proofs of theorems.
        \item[] Appedix C: Statistical tests based on different scores.
        \item[] Appedix D: Additional plots with synthetic data for tree comparisons.
        \item[] \revision{Appedix E: Finding the true splits.}
        \item[] Appedix F: Real data descriptions.
    \end{enumerate}

\appendix
\counterwithin{figure}{section}
\counterwithin{equation}{section}
\renewcommand\thefigure{\thesection.\arabic{figure}} 
\renewcommand\theequation{\thesection.\arabic{equation}} 

\section{List of the existing pre-pruning algorithms}
\label{suppA}

Pre-pruning is a cost-effective approach to pruning, and its common approaches include:
\begin{enumerate}
    \item finding the smallest tree that is within one standard deviation from the numerically optimal value, 
    \item penalizing the overall error with the number of terminal nodes with a unit cost~\citep{breiman1984classification}, i.e., $S_\kappa(F,G) = S(F,G)+m \times \kappa$, where $m$ is the number of terminal nodes and $\kappa= \mcO(-\frac{p}{n}\log(\frac{p}{n}))$~\citep{klusowski2020sparse},
    \item using p-values to bring comparisons in the same scale and stopping the tree growth with a threshold to statistical significance, a.k.a, conditional inference trees ~\citep{hothorn2006unbiased} -- p-values will increase with the size of the tree,
    \item comparing expected error reductions measured by weighted standard deviation of responses~\citep{zeileis2008model} with a threshold, i.e.,  $\sigma-\sum_{t=1}^m\frac{n_t}{n}\sigma_{t}\leq \kappa$, where  $\sigma=\sqrt{\sum_{i\in \mcJ}(y_i-\ybar_{\mcJ})^2}/\sqrt{n-1}$ and $\sigma_{t}=\sqrt{\sum_{i\in \mcJ_t}(y_i-\ybar_{\mcJ_t})^2}/\sqrt{n_t-1}$, 
    \item finding inadequate subtrees by their effect in the \emph{adjusted error rate}, which is computed by $\frac{n+m}{n-m}\sum_{i\in \mcJ}|y_i-\hat{y}_i|$ with $\hat{y}$ being the predictive value of each point, and 
    \item recursive shrinking~\citep{hastie1990shrinking}, which is to combine predictions at the node and a model that would have been used at the root node with $(n_t\hat{y}+c\ybar_{\mcJ})/(n_t+c)$, where 
    $c$ is a constant with a default $15$.
\end{enumerate}

\section{Theorem Proofs}
\label{suppProofs}
\revision{
\subsection{Proof of Theorem~\ref{thm:monotone}}
\begin{proof}
The proof relies only on the basic feature of proper scoring rules. Following the property of proper scoring rules~\eqref{eq:proper}, we can write 
\begin{align}
 \sum_{i \in \mcL_{t}(k,s)} S&\left(\Fhat_{\mcL_{t}(k,s)},y_i\right) + \sum_{i \in \mcR_{t}(k,s)} S\left(\Fhat_{\mcR_{t}(k,s)},y_i\right)\nonumber\\ 
 = & |\mcL_t(k,s)| \text{ES}(\widehat{F}_{\mcL_t(k,s)},\widehat{F}_{\mcL_t(k,s)}) +|\mcR_t(k,s)| \text{ES}(\widehat{F}_{\mcR_t(k,s)},\widehat{F}_{\mcR_t(k,s)})\nonumber\\
 \leq & |\mcL_t(k,s)| \text{ES}(\widehat{F}_{\mcJ_t},\widehat{F}_{\mcL_{t}(k,s)}) + |\mcR_t(k,s)| \text{ES}(\widehat{F}_{\mcJ_t},\widehat{F}_{\mcR_{t}(k,s)}) \nonumber\\
 =& |\mcL_t(k,s)| \frac{1}{|\mcL_t(k,s)| } \sum_{i \in \mcL_{t}(k,s)} S\left(\Fhat_{\mcJ_{t}},y_i\right) + |\mcR_t(k,s)| \frac{1}{|\mcR_t(k,s)|} \sum_{i \in \mcR_{t}(k,s)} S\left(\Fhat_{\mcJ_{t}},y_i\right) \nonumber \\
 =& \sum_{i \in \mcJ_t} S(\widehat{F}_{\mcJ_t},y_i) , \nonumber
\end{align}
\noindent where the first and second equality use the definition of $\text{ES}(\cdot,\cdot)$.
\end{proof} 
\subsection{Proof of Theorem \ref{thm:consist_gen}}
\begin{proof} For the readability of this proof, we make a number of changes in the notation.  Let us denote ``side $1$ of the data split'' be all $y_i$ such that $x_i \in \hat A_n$ and ``side $2$ of the data split'' be all $y_i$ such that $x_i \in \hat A^{c}_n$, i.e., $x_i \notin \hat A_n$. Denote $A^{*}_n$ as the oracle split choice. We adopt a similar definition for the oracle split. Also, define 
 $$F_{ij} = \begin{cases} \text{ distribution of } y \text{ given } x \in \hat A_n \cap A^{*}_n & \text{ if } i =1, j=1 \\
\text{ distribution of } y \text{ given } x \in \hat A^{c}_n \cap A^{*}_n & \text{ if } i =1, j=2 \\
\text{ distribution of } y \text{ given } x \in \hat A_n \cap A^{*c}_n & \text{ if } i =2, j=1 \\
\text{ distribution of } y \text{ given } x \in \hat A^{c}_n \cap A^{*c}_n & \text{ if } i =2, j=2,
\end{cases}$$ the associated subregion's true distribution, and $$\gamma_{ij} = \begin{cases} \Pr\left(x \in \hat A_n \cap A^{*}_n\right) & \text{ if } i =1, j=1 \\
 \Pr\left(x \in \hat A^{c}_n \cap A^{*}_n\right) & \text{ if } i =1, j=2 \\
 \Pr\left(x \in \hat A_n \cap A^{*c}_n\right) & \text{ if } i =2, j=1 \\
 \Pr\left(x \in \hat A^{c}_n \cap A^{*c}_n\right) & \text{ if } i =2, j=2 , 
\end{cases}$$ the associated subregion's true probability. Furthermore, define  $\lambda:=\Pr(x\in \hat A_n)$ and $\pi:=\Pr(x\in A^{*}_n)$ that using the subregion probabilities can be computed as $\lambda=\gamma_{11}+\gamma_{21}$ and $\pi=\gamma_{11}+\gamma_{12}$. The true distributions on side $1$ of the data split is
\[F_1 = \frac{\gamma_{11}}{\gamma_{11}+\gamma_{21}} F_{11} + \frac{\gamma_{21}}{\gamma_{11}+\gamma_{21}} F_{21}, \text{ and on side } 2,
F_2 = \frac{\gamma_{12}}{\gamma_{12}+\gamma_{22}} F_{12} + \frac{\gamma_{22}}{\gamma_{12}+\gamma_{22}} F_{22}.\] Note, $F_1 = F_{\mcL(\hat A_n;\infty)}$ and $F_2 = F_{\mcR(\hat A_n;\infty)}$. Similarly, for the oracle split, the true distribution of points on side $1$ is
\[F_1^* = \frac{\gamma_{11}}{\gamma_{11}+\gamma_{12}} F_{11} + \frac{\gamma_{12}}{\gamma_{11}+\gamma_{12}} F_{12}, \text{ and on side } 2, F_2^* = \frac{\gamma_{21}}{\gamma_{21}+\gamma_{22}} F_{21} + \frac{\gamma_{22}}{\gamma_{21}+\gamma_{22}} F_{22}.\] Note, $F_1^* = F_{\mcL(A^*_n;\infty)}$ and $F_2^* = F_{\mcR(A^*_n;\infty)}$. 

Let $n_{ij}$ be the number of points on side $i$ of the oracle split and $j$ on the data-based split. Let $\Fhat_{ij}$ be the empirical distribution of the $n_{ij}$ data points on side $i$ of the oracle split and $j$ of the data-based split. Define the fraction assigned to side $1$ in the data split as 
$$\hat{\lambda} = \widehat{\Pr}(x\in \hat A_n) = \frac{n_{11} + n_{21}}{n},$$ the fraction assigned to side $1$ in the oracle split as
$$\hat{\pi} = \widehat{\Pr}(x\in A^{*}_n) = \frac{n_{11}+n_{12}}{n}.$$

The empirical distribution of points on side $1$ of the data split is
\[\Fhat_1 = \frac{n_{11}}{n_{11}+n_{21}} \Fhat_{11} + \frac{n_{21}}{n_{11}+n_{21}} \Fhat_{21}, \text{ and on side } 2,
\Fhat_2 = \frac{n_{12}}{n_{12}+n_{22}} \Fhat_{12} + \frac{n_{22}}{n_{12}+n_{22}} \Fhat_{22}.\]  Note, $\Fhat_1 = \Fhat_{\mcL(\hat A_n;n)}$ and $\Fhat_2 = \Fhat_{\mcR(\hat A_n;n)}$. For the oracle split, the empirical distribution of points on side $1$ is
\[\Fhat_1^* = \frac{n_{11}}{n_{11}+n_{12}} \Fhat_{11} + \frac{n_{12}}{n_{11}+n_{12}} \Fhat_{12}, \text{ and on side } 2, \Fhat_2^* = \frac{n_{21}}{n_{21}+n_{22}} \Fhat_{21} + \frac{n_{22}}{n_{21}+n_{22}} \Fhat_{22}.\] Note, $\Fhat_1^* = \Fhat_{\mcL(A^*_n;n)}$ and $\Fhat_2^* = \Fhat_{\mcR(A^*_n;n)}$. 
By the definition of the data split that finds the smallest split for the data, we get
\begin{equation}
    \hat{\lambda} \ES(\Fhat_1,\Fhat_1)+(1-\hat{\lambda}) \ES(\Fhat_2,\Fhat_2) \leq \hat{\pi} \ES(\Fhat_1^*,\Fhat_1^*) +(1-\hat{\pi}) \ES(\Fhat_2^*,\Fhat_2^*).\label{eq:empirical-result} 
\end{equation}

Using the mixture distributions structure and observing that for some arbitrary probability distributions $H,F,$ and $G$
\begin{align*}
    \ES\left(H,\frac{n_{F}}{n_{F}+n_{G}}F+\frac{n_{G}}{n_{F}+n_{G}}G\right)&=\mbE_{y\sim \frac{n_{F}}{n_{F}+n_{G}}F+\frac{n_{G}}{n_{F}+n_{G}}G}[S(H,y)]\\
    &=\int S(H,y) d\left(\frac{n_{F}}{n_{F}+n_{G}}F+\frac{n_{G}}{n_{F}+n_{G}}G\right)(y)\\
    &=\frac{n_{F}}{n_{F}+n_{G}}\ES(H,F)+\frac{n_{G}}{n_{F}+n_{G}}\ES(H,G),
\end{align*}
we find that inequality~\eqref{eq:empirical-result} then translates to
\begin{align} 
\frac{n_{11}}{n} \ES(\Fhat_1,\Fhat_{11}) + &\frac{n_{12}}{n} \ES(\Fhat_2,\Fhat_{12}) +\frac{n_{21}}{n} \ES(\Fhat_1,\Fhat_{21}) + \frac{n_{22}}{n} \ES(\Fhat_2,\Fhat_{22}) \label{eq:empirical-result2} \\
 \leq& \frac{n_{11}}{n} \ES(\Fhat_1^*,\Fhat_{11})+ \frac{n_{12}}{n} \ES(\Fhat_1^*,\Fhat_{12})+\frac{n_{21}}{n} \ES(\Fhat_2^*,\Fhat_{21}) + \frac{n_{22}}{n} \ES(\Fhat_2^*,\Fhat_{22}). \nonumber
\end{align}

We next show, fixing $\Fhat_i$s and $\Fhat^*_i$s (and the tree that is trained with the available data), the expected score of the empirical splits will approach the smallest possible score (that would be obtained with the oracle splits). That is, we show as the size of the testing data (second argument of the expected score terms $\ES(\Fhat_j,\Fhat_{ij})$) grows large, the left-hand side of inequality \eqref{eq:empirical-result2} will converge in probability to $g(\hat A_n)$ and similarly the right-hand side will converge in probability to $g(A_n^*)$ or $g^*$.
To that end, we first write
\begin{align*}
    g(\hat A_n) &=\lambda \ES\left(\Fhat_1,\frac{\gamma_{11}}{\gamma_{11}+\gamma_{21}} F_{11} + \frac{\gamma_{21}}{\gamma_{11}+\gamma_{21}} F_{21}\right)+(1-\lambda) \ES\left(\Fhat_2,\frac{\gamma_{12}}{\gamma_{12}+\gamma_{22}} F_{12} + \frac{\gamma_{22}}{\gamma_{12}+\gamma_{22}} F_{22}\right)\\
    &=\gamma_{11} \ES(\Fhat_1,F_{11})+ \gamma_{12}\ES(\Fhat_2,F_{12}) + \gamma_{21} \ES(\Fhat_1,F_{21}) + \gamma_{22} \ES(\Fhat_2,F_{22}),
\end{align*} and 
\begin{align*}
    g(A^{*}_n)&= \pi \ES\left(\Fhat_1^*, \frac{\gamma_{11}}{\gamma_{11}+\gamma_{12}} F_{11}+ \frac{\gamma_{12}}{\gamma_{11}+\gamma_{12}} F_{12}\right)+(1-\pi) \ES\left(\Fhat_2^*, \frac{\gamma_{21}}{\gamma_{21}+\gamma_{22}} F_{21} + \frac{\gamma_{22}}{\gamma_{21}+\gamma_{22}} F_{22}\right)\\
    &=\gamma_{11} \ES(\Fhat_1^*,F_{11})+\gamma_{12} \ES(\Fhat_1^*,F_{12}) +\gamma_{21} \ES(\Fhat_2^*,F_{21}) + \gamma_{22} \ES(\Fhat_2^*,F_{22}).
\end{align*}

By construction, either $n_{ij}=0$ if no chance of an $x$ being in that intersection and $\gamma_{ij}=0$, or $n_{ij}$ goes to $\infty$ and $n_{ij}/n \rightarrow \gamma_{ij}$ in probability. For all those subregions with $\gamma_{ij} > 0$, we require that
\begin{equation}
    \ES(\Fhat_j,\Fhat_{ij})\inP \ES(\Fhat_j,F_{ij}), \text{ and }\ \ES(\Fhat_i^*,\Fhat_{ij})\inP  \ES(\Fhat_i^*,F_{ij}).\label{eq:required-conditions}
\end{equation}
These conditions are met directly from our theorem postulates. Thus we can conclude that in our cases for all $\epsilon > 0$,
\begin{align} 
\Pr \left( g(\hat A_n) - g^{*} >\epsilon \right) \rightarrow 0, \nonumber
\end{align}
as $n\to\infty$ (again keeping the first argument of $\ES(\cdot,\cdot)$ terms fixed). Since $g(\hat A_n) \geq g^{*}$ by definition, the desired result follows. For the general case and arbitrary choice of a proper score for building trees, it is clear that the condition \eqref{eq:required-conditions} is exactly what is needed to guarantee this result. In particular scoring cases that were in focus in this article, the next corollary will clarify what they translate to. 
\end{proof}
\subsection{Proof of Corollary \ref{thm:consist_CRPS}}
\begin{proof}
    We show that with the finite second moment of the targets in all potential subregions, the postulate of Theorem \ref{thm:consist_gen} will hold and hence, the result of $g(\hat A_n)\inP g^*$ will be proven. Without loss of generality, we show this using $i=2$ and $j=1$ for each scoring rule. For CRPS, our moment condition leads to 
\begin{align*}
    \ES(\Fhat_1,\Fhat_{21}) - \ES(\Fhat_1,F_{21}) & = \frac{1}{n_{21}}\sum_{i:x_i \in A^{*c}_n \cap \hat A_n}S(\Fhat_1,y_i) - \mbE_{y\sim F_{21}}[S(\Fhat_1,y)]\\
    &=\frac{1}{n_{21}}\sum_{i:x_i \in A^{*c}_n \cap \hat A_n} \mbE_{y\sim \Fhat_1}|z-y_i| -\frac{1}{2} \mbE_{z,z'\sim \Fhat_1}|z-z'| \\
    &\quad\quad -\left(\mbE_{y\sim F_{21}} \mbE_{z\sim \Fhat_1}|z-y| - \frac{1}{2} \mbE_{z,z'\sim \Fhat_1}|z-z'| \right)\\
    & = \mbE_{y\sim \Fhat_1}\left[\frac{1}{n_{21}}\sum_{i:x_i \in A^{*c}_n \cap \hat A_n}|z-y_i|-\mbE_{y\sim F_{21}}|z-y|\right]
\end{align*}where in the last equality the inner- and outer-expectations are interchanged by Fubini's theorem. To complete the proof for CRPS, we invoke the empirical theory that implies $\mbE_{y\sim \Fhat_{21}}|z-y|\inP \mbE_{y\sim F_{21}}|z-y|$. This then leads to the term inside the expectation in the last equality to converge to 0 in probability.
    

Next, for DSS we leverage the moment condition to write 
\begin{align*}
    \ES(\Fhat_1,\Fhat_{21}) - \ES(\Fhat_1,F_{21}) & = \mbE_{y\sim \Fhat_{21}}\left[\frac{(\hat{\mu}_1-y)^2}{\hat{\sigma}_{1}^2}+\ln{\hat{\sigma}_{1}^2}\right]-\mbE_{y\sim F_{21}}\left[\frac{(\hat{\mu}_{1}-y)^2}{\sigma^2_{\Fhat_1}}+\ln{\hat{\sigma}_{1}^2}\right]\\
    &=\frac{1}{\hat{\sigma}_{1}^2}\left(\frac{1}{n_{21}}\sum_{x_j \in A^{*c}_n \cap \hat A_n} (\hat{\mu}_{1}-y_i)^2-\mbE_{y\sim F_{21}}[(\hat{\mu}_{1}-y)^2]\right),
\end{align*} where $\hat{\mu}_{1}$ and $\hat{\sigma}_{1}^2$ is the mean and variance of $\Fhat_{1}$.
As long as $y$ has a finite second moment for any arbitrary subregion, invoking the typical law of large numbers for continuous functions tends the term in parenthesis to drop to zero in probability, proving the results for this scoring rule. 

Finally, for the IS1 score, we use the definition to write 
\begin{align*}
    \ES(\Fhat_1,\Fhat_{21}) - \ES(\Fhat_1,F_{21}) & = \frac{1}{n_{21}}\sum_{x_j \in A^{*c}_n \cap \hat A_n}\left(y_i-\hat{q}_{1} (1-\alpha)\right)^+ - \mbE_{y\sim F_{21}}\left(y-\hat{q}_{1} (1-\alpha)\right)^+\\
    &= \frac{1}{n_{21}}\sum_{x_j \in A^{*c}_n \cap \hat A_n,  y_i \geq \hat{q}_{1}(1-\alpha)}\left|y_i-\hat{q}_{1} (1-\alpha)\right| \\
    & \quad \quad\quad - \mbE_{y\sim F_{21}}[|y-\hat{q}_{1} (1-\alpha)|\mbI(y \geq \hat{q}_{1}(1-\alpha))],
\end{align*} where $\hat{q}_{1} (1-\alpha)$ is the $1-\alpha$ quantile of $\Fhat_{1}$, $\mbI(\cdot)$ represent an indicator function, and an operator $(a)^+:=\max\{a,0\}$ for some $a$. 
Therefore, as long as the subregions' tail probabilities at $\hat{q}_{i} (1-\alpha)$ are consistent, which is ensured by our condition on the cumulative distribution function, the right-hand side of the above converges to zero in probability and hence the result stands. 

Note, for all the cases above, the same arguments can be made to show $\ES(\Fhat_1^*,\Fhat_{21}) - \ES(\Fhat_1,F_{21})$ will converge to zero in probability, completing the proof.
\end{proof}
}

\section{Statistical tests based on different scores}
\label{suppF}

\revision{We perform the following Hypothesis tests:
\begin{itemize}
    \item[$H_0^{\text{SSE}}$:] Trees built with non-SSE scores result in worse SSE than trees built with SSE.
    \item[$H_0^{\text{CRPS}}$:] Trees built with non-CRPS scores result in worse CRPS than trees built with CRPS.
    \item[$H_0^{\text{DSS}}$:] Trees built with non-DSS scores result in worse DSS than trees built with DSS.
    \item[$H_0^{\text{IS1}}$:] Trees built with non-IS1 scores result in worse IS1 than trees built with IS1.
\end{itemize}
Failing to reject $H_0$ would imply that when the goal of prediction is to minimize loss measured in a particular scoring rule, trees trained that same scoring rule as the splitting criteria will yeild the best predictions. Specifically, for $H_0^{\text{SSE}}$, the mean SSE value from non-SSE-based trees is higher than the mean SSE value from SSE-based trees. } We compute the out-of-sample p-values using a paired-t-test (pairing for each replication) for all \text{Build} and \text{Eval}$\neq$\text{Build} scores. We reject the hypothesis \revision{corresponding to each \text{Eval} score when the computed p-value is smaller than 0.05. } 

\renewcommand{\arraystretch}{0.75}
\begin{table}[H]
\caption{\revision{Out-of-sample p-values smaller than $0.05$ (signified by bold fonts) implies the \text{Eval}-based tree leads to higher (worse) \text{Eval} score than the \text{Build}-based tree. $\kappa=0$ means no pruning and $\kappa=0.1$ means mild pruning.}
}
\footnotesize
\centering
\begin{tabular}{ |c|c|c|c c c c | c c c c| } 
\cline{4-11}
\multicolumn{3}{c|}{} & \multicolumn{4}{c|}{$\kappa=0.0$} & \multicolumn{4}{c|}{$\kappa=0.1$} \\
  \cline{2-11}
\multicolumn{1}{c|}{}&\text{Eval} score
& \diagbox{$n$}{\text{Build} score}  & SSE & CRPS & DSS & IS1 & SSE & CRPS & DSS & IS1 \\ \cline{2-11} \hline
 	\parbox[t]{2mm}{\multirow{16}{*}{\rotatebox[origin=c]{90}{Easy Dataset}}}
 	&\multirow{4}{*}{SSE} & 200 & -- & 0.51 & 1.00& 1.00 & -- & \textbf{0.02} & \textbf{0.02} & 0.52\\ 
 	 && 400 & -- & 0.09 & 0.06 & 0.80 & --  & 0.05 & \textbf{0.02} & 0.75  \\
 	 && 800 & -- & 0.72 & 1.00 & 0.92& --  & 0.10& 0.19 & 0.79 \\  
 	 && 1600 & -- & 0.33 & 1.00& 0.73 & -- & 0.08 & 0.17 & 0.90\\ \cline{2-11}
 	&\multirow{4}{*}{CRPS} & 200 & 0.83 & -- & 1.00 & 1.00 & 0.98 & -- & 0.07 & 1.00 \\ 
 	 && 400 & 0.99 & -- & 0.45 & 0.92& 0.95 & -- & \textbf{0.01} & 1.00 \\ 
 	 && 800 & 0.59 & -- & 1.00 & 0.94& 0.90 & -- & 0.99 & 1.00 \\ 
 	 && 1600 & 0.84 & -- & 1.00 & 0.90 & 0.92 & -- & 0.99 & 1.00 \\ \cline{2-11}
 	&\multirow{4}{*}{DSS} & 200 & 0.65 & 0.58 & -- & 1.00 & 1.00 & 1.00 & -- & 1.00 \\ 
 	 && 400 & 0.71 & 0.29 & -- & 1.00 & 1.00 & 1.00 & -- & 1.00 \\ 
 	 && 800 & 0.06 & \textbf{0.04} & -- & 0.98 & 0.99 & 0.96 & -- & 1.00 \\ 
 	 && 1600 & 0.37 & 0.21 & -- & 0.95 & 1.00 & 1.00 & -- & 1.00 \\ \cline{2-11} 
 	&\multirow{4}{*}{IS1} & 200 & 1.00 & 1.00 & 1.00 & -- & 0.99 & 1.00 & 1.00 & -- \\ 
 	 && 400 & 1.00 & 1.00 & 1.00 & -- & 0.97 & 0.98 & 0.65 & -- \\ 
 	 && 800 & 1.00 & 1.00 & 1.00 & -- & 0.98 & 1.00 & 1.00 & -- \\ 
 	 && 1600 & 1.00 & 1.00 & 1.00 & -- & 0.97 & 0.97 & 0.99 & -- \\ \cline{2-11}
\hline
    \parbox[t]{2mm}{\multirow{16}{*}{\rotatebox[origin=c]{90}{Hard Dataset}}}
 	&\multirow{4}{*}{SSE} & 200 & -- & 0.25 & \textbf{0.01} & \textbf{0.01} & -- & 0.22 & \textbf{0.00} & \textbf{0.00} \\ 
 	 && 400 & -- & \textbf{0.00} & \textbf{0.00} & \textbf{0.00} & -- & \textbf{0.00} & \textbf{0.00} & \textbf{0.00} \\ 
 	 && 800 & -- & \textbf{0.00} & \textbf{0.00} & \textbf{0.00} & -- & \textbf{0.00} & \textbf{0.00} & \textbf{0.00} \\  
 	 && 1600 & -- & \textbf{0.00} & \textbf{0.00} & \textbf{0.00} & -- & \textbf{0.00} & \textbf{0.00} & \textbf{0.00} \\ \cline{2-11}
 	&\multirow{4}{*}{CRPS} & 200 & 0.89 & -- & 0.15 & \textbf{0.01} & 0.95 & -- & 0.16 & \textbf{0.00} \\ 
 	 && 400 & 1.00 & -- & 0.64 & 0.22& 1.00 & -- & 0.47 & 0.15\\ 
 	 && 800 & 1.00 & -- & 0.79 & 0.15 & 1.00 & -- & 0.77 & 0.09 \\ 
 	 && 1600 & 1.00 & -- & 0.41 & \textbf{0.02} & 1.00 & -- & 0.46 & \textbf{0.04}\\ \cline{2-11}
 	&\multirow{4}{*}{DSS} & 200 & 0.98 & 0.15 & -- & 0.07 & 0.99 & 0.37 & -- & 0.14\\ 
 	 && 400 & 0.92 & \textbf{0.03} & -- & \textbf{0.02} & 1.00 & 0.17 & -- & 0.07\\ 
 	 && 800 & 0.97 & 0.07 & -- & \textbf{0.03} & 1.00 & 0.29 & -- & 0.13\\ 
 	 && 1600 & 1.00 & 0.08 & -- & 0.05 & 1.00 & 0.64 & -- & 0.61\\ \cline{2-11} 
 	&\multirow{4}{*}{IS1} & 200 & 0.97 & 0.26 & 0.14 & -- & 0.99 & 0.33 & 0.40 & -- \\ 
 	 && 400 & 0.88 & 0.07 & \textbf{0.04} & -- & 0.98 & 0.08 & 0.16 & -- \\ 
 	 && 800 & 0.89 & \textbf{0.03} & \textbf{0.02} & -- & 0.98 & 0.07 & 0.80 & -- \\ 
 	 && 1600 & 1.00 & 0.13 & 0.06 & -- & 1.00 & 0.12 & 0.47 & -- \\ \cline{2-11}
\hline

  \end{tabular}\label{tab:hypotest}
\end{table}

Our analysis signifies the performance of each score-based tree when compared with other trees by computing the p-values for pruning parameters $\kappa=0$ and 0.1. The general trend in Table~\ref{tab:hypotest}, especially for the easy dataset, is that \text{Eval} trees are better than \text{Build} trees when compared in \text{Eval} score. 

Table~\ref{tab:hypotest} also shows that the pruning parameter particularly makes a difference for the consistency of DSS and IS1 trees. This effect is more vivid in the hard dataset, where several cases that would be rejected when $\kappa=0.0$ will no longer be rejected when  $\kappa=0.1$. It also shows that the small pruning may not be sufficient for the consistency of the SSE or CRPS trees when compared with other score trees but with the same pruning value. To explore the best possible performance of each score based-tree, we next choose a tuned pruning parameter to see whether the poor performance of SSE 
especially in the hard dataset can be remedied with better pruning.

\section{Finding the true splits}\label{suppH}

\begin{table}[H]
\centering
\footnotesize
\caption{Detection of true splits with optimal pruning $\kappa^{*}(\text{\text{Eval}})$ across 30 replicates.}\label{tab:split-interpret}
\begin{tabular}{|c|ccc|c|}
\hline
\multicolumn{5}{|c|}{{\bf Hard Dataset with $n=1600$}}\\ \hline
 \diagbox{\text{Eval}}{true split} & $-0.5$ & $0$ & $0.5$  & average \# of incorrect splits \\ \hline
 $\text{SSE}$ & 10\% & 70\% & 70\% & 3.9 \\ \hline
 $\text{CRPS}$ & 77\% & 87\% & 90\% & 1.7 \\ \hline
 $\text{DSS}$ & 73\% & 90\% & 67\% & 4.0 \\ \hline
 $\text{IS1}$ & 73\% & 93\% & 83\% & 2.5\\ \hline
\end{tabular}
\end{table}
We compare the trees not in terms of their resulting scores but with respect to the true tree structure that we have designed (See Table~\ref{tab:synth-data}). For ease of exposition, we focus on the hard dataset. Note, in the hard dataset the behavior of each true partition of the data starts to look more different only in the third moment and looks quite similar in the first and second moments of each partition. Also, the behavior of data is more homogeneous for $x<0$ and $x>0$ but the further splits are harder to catch. We report which of the scoring rules is more successful in recovering the true partitions. For that, using each scoring rule, we check all 30 resulting tree structures built with the optimal pruning threshold. We count the number of times splitting at $-0.5,0,$ and $0.5$ are discovered.

We also count the number of incorrect splits in each of the scores' resulting 30 trees. This is because we have allowed the trees to be of depth 4, i.e., a total of 15 split values, and even the tuned trees that may be asymmetric can return more than 3 split points. We report the average number of incorrect splits at optimal pruning in the last column of Table~\ref{tab:split-interpret}. This value also shows that the SSE and DSS trees tend to find more incorrect splits in the data (3.9 and 4), while CRPS trees find the fewest incorrect splits (1.7).
Regarding the correct splits, the average mean and variance of data when $x< 0$ and $x> 0$ are still easier to distinguish and all scores find the $x<0$ split in more than half of the runs. However, the other two split values are harder to recover. While non-SSE trees tend to do better in finding them, we also observe that IS1 is more successful than DSS in recovering these more challenging split values, and CRPS is more successful than both.

\section{Real data description}
\label{supprealdata}

\begin{description}
\item[Ethiopian yield dataset:] The yield data were sourced from the Ethiopian Central Statistics Agency’s Annual Agricultural Sample Surveys. Data were spatially referenced and then mapped to climatic data sourced from GDAS (temperature) and CHIRPS (precipitation) that was bias-corrected and downscaled using MicroMet formulations in the NASA Land Information System. Soil data were sourced from ISIRC.

\item[Divvy bikeshare dataset: ] Lyft Bikes and Scooters, LLC (“Bikeshare”) operates the City of Chicago’s (“City”) Divvy bicycle sharing service. Historical trip data including Trip start day and time,
Trip end day and time,
Trip start station,
Trip end station,
Rider type (Member, Single Ride, and Day Pass).
The data has been processed to remove trips that are taken by staff as they service and inspect the system; and any trips that were below 60 seconds in length (potentially false starts or users trying to re-dock a bike to ensure it was secure).

\end{description}
%
%
%